\documentstyle[12pt]{article} 

\begin{document}

\bigskip\bigskip\bigskip

\begin{center}
{\Large \bf Universal Superpositions of Coherent States \\[2mm]
and Self-Similar Potentials}\\[10mm]
Vyacheslav Spiridonov%
\footnote{On leave of absence from the Institute for Nuclear Research,
Russian Academy of Sciences, Moscow, Russia}\\ [2mm]
{\it Centre de Recherches Math\'ematiques, Universit\'e de Montr\'eal,\\
C.P. 6128, succ. Centre-ville, Montr\'eal, Qu\'ebec, H3C 3J7, Canada\\
e-mail: spiridonov@lps.umontreal.ca }\\[7mm]

\end{center}

\vspace{2mm}

\begin{abstract}
A variety of coherent states of the harmonic oscillator  is considered. 
It is formed by a particular superposition of canonical 
coherent states. In the simplest case, these superpositions are eigenfunctions 
of the annihilation operator $A=P(d/dx+x)/\sqrt2$, 
where $P$ is the parity operator. 
Such $A$ arises naturally in the $q\to -1$ limit for a symmetry 
operator of a specific self-similar potential obeying the $q$-Weyl 
algebra, $AA^\dagger-q^2A^\dagger A=1$. Coherent states 
for this and other reflectionless potentials whose discrete spectra consist
of $N$ geometric series are analyzed. In the harmonic oscillator limit
the surviving part of these states takes the form of 
orthonormal superpositions of $N$ 
canonical coherent states $|\epsilon^k\alpha\rangle$, $k=0, 1, \dots, N-1$, 
where $\epsilon$ is a primitive $N${th} root of unity, $\epsilon^N=1$.
A class of $q$-coherent states related to the bilateral
$q$-hypergeometric series and Ramanujan type integrals
is described. It includes a curious set of
coherent states of the free nonrelativistic
particle which is interpreted as a $q$-algebraic system without
discrete spectrum. A special 
degenerate form of the symmetry algebras of self-similar potentials
is found to provide a natural $q$-analog of the Floquet theory.
Some properties of the factorization method, which is used throughout 
the paper, are discussed from the differential Galois theory point of view.

\medskip
\noindent
PACS numbers: 03.65.-w,  42.50.-p, 02.30.-f

\end{abstract}

\begin{center}
\bf Published in Phys. Rev. A52, no. 3 (1995) 1909-1935.
\end{center}

\newpage

\tableofcontents
\bigskip

\newpage

\section{Introduction}
\setcounter{equation} 0

Replacement of commuting coordinate and momentum variables of a classical
point particle by the operators $x$ and $p$ satisfying the Heisenberg
commutation relation,
\begin{equation}
[x,p] \equiv xp-px=i\hbar,
\label{ccr}
\end{equation}
endows this particle with wave characteristics. 
According to the original definition, coherent states are the states 
in which corpuscular properties of a quantum particle are seen best of all.
For the harmonic oscillator,  such a qualitative motivation
happens to be supported by the rich group-theoretical content built in the 
structure of coherent states, 
so that the symmetry approach provides their alternative description. 
As a result, there have appeared several different quantitative definitions of
coherent states which are equivalent only for the harmonic 
oscillator case. Independently
of this non-uniqueness of definition, from both the physical and 
mathematical points of view, coherent states of quantum mechanics are 
fascinating objects  having useful 
applications in many fields \cite{coh}-\cite{gilmore}.

In the present paper we discuss coherent states associated with a 
specific class of one-dimensional 
Schr\"odinger operator potentials found in \cite{shabat}-\cite{s2}. 
The general class of these {\it self-similar} potentials
is defined with the help of $q$-periodic closure \cite{s1,s2} of the 
dressing chain, or the chain of Darboux transformations. The latter
transformations are known to be closely related to the 
factorization method \cite{infeld,miller}.
Symmetries of the self-similar potentials are
described by some polynomial operator
algebras of order $N$ ($N$ is the period of closure)
which play the role of spectrum generating algebras. 
For $N=1, \; 2$ 
these algebras coincide with known $q$-analogs of 
the bosonic oscillator and $su(1,1)$ algebras \cite{s1}. 
The {\it $q$-coherent states}, defined as eigenfunctions of symmetry operators
which lower the energy, have many interesting properties. 
In particular, the algebras depend on the parameter $q^2$, so that 
$q=\pm1$ coherent states seem to be equivalent; but this is not so.
For the $q$-Weyl algebra system, the limit $q\to -1$ exists
only when the potential is symmetric, and then the corresponding coherent 
states are described by a particular superposition of canonical
coherent states. This superposition has a universal form,  it represents an 
example of the Titulaer-Glauber coherent states \cite{titulaer}
which was constructed in \cite{yurke} from a different idea.
Its multimode oscillator analog defines 
the particular entangled coherent states which have a 
phase difference equal to $\pi/2$. The name {\it parity coherent states}
is suggested for these and other more general two-term
superpositions of coherent states for which the parity operator plays a
crucial role in the definition.

For the limiting values of parameters corresponding to the harmonic oscillator
potential, the raising and lowering operators of the self-similar 
potentials' symmetry algebras become equal to powers of 
bosonic creation and annihilation operators. In this limit, part of the 
$q$-coherent states degenerate into orthonormal superpositions of 
$N$ canonical coherent states $|\epsilon^k\alpha\rangle, k=0, 1,\dots, N-1,$
where $\epsilon$ is a primitive $N${th} root of unity, $\epsilon^N=1$. 
These are natural generalizations of even and odd coherent 
states \cite{dodonov}. Note that there are $q$-coherent states which
do not survive in this limit. In another degenerate limit, when the potential
vanishes but $q$ remains arbitrary, one gets a unique set of
non-trivial coherent states of the free particle system.

It is interesting that symmetry operators of the self-similar 
potentials are defined with the help of a scaling operator -- 
a special form of the squeezing operator.
Due to this fact, wave functions of these systems 
resemble wavelets. From a degenerate form of the symmetry algebra,
when pure dilatation by a fixed number $q$ becomes physical  
symmetry of the Schr\"odinger equation, one finds a natural 
$q$-analog of the Floquet theory.

Despite their simplicity, group-theoretical roots of superpositions of 
canonical coherent states comprise an intrinsic possibility for
construction of complicated physical systems whose coherent states
share common properties with these superpositions.
The author considers the description of the relationship between self-similar
potentials and superpositions of coherent states 
as an important physical result of this paper. However,
a wider aim of the work is to discuss exactly solvable
potentials and their coherent states
from the general functional-analytic point of view on the basis of
old and new examples. 

The paper is organized as follows. Before moving to the analysis of 
complicated situations, at the end of this section we give
a brief account of the canonical coherent states of the 
harmonic oscillator. In section 2 we construct non-standard
coherent states of the same system and consider their relation to the
Titulaer-Glauber coherent states. 
In section 3 we discuss a universality of the derived 
superpositions of coherent states. An interesting set of coherent states
of the free particle determined by the pantograph equation and its 
generalizations is considered in section 4.
The simplest potentials with $q$-deformed
symmetry algebras and non-trivial discrete spectra 
are described in section 5, where some properties
of the associated coherent states are analyzed. In the next, $6${th} 
section we present a general hierarchy of Schr\"odinger operators whose
discrete spectra consist of $N$ geometric series generated by the
specific polynomial quantum algebras. 
Two particular systems arising from $N=2, \; q=-1$ and 
$N=3, \; q=1$ closures of the dressing chain are considered
in section 7. In section 8, a $q$-analog of the Floquet theory is outlined.
Section 9 contains a discussion of integrable potentials 
and coherent states from the differential (``quantum") Galois theory point 
of view. Some concluding remarks are given in section 10. 
The paper has a formal character; we consider mostly theoretical
aspects of the chosen (stationary) 
systems rather than their possible experimental implementations.
Present analysis of coherent states for the self-similar potentials arose
from the investigation of $q$-oscillator algebra at roots of unity 
performed in \cite{skorik}.
The results of this work have been reported by the author 
in \cite{can,skorik}, and 
a part of them has been published in \cite{s3}. 

In practical applications it is convenient to use the coordinate 
representation of (\ref{ccr}), where $p=-i\hbar d/dx$. 
For simplicity we use the system of units where
Plank's constant $\hbar$ is equal to 1.
In terms of the ladder operators $a^\dagger, a$:
\begin{equation}
a=\frac{1}{\sqrt2}\left(\frac{d}{dx}+x\right), \qquad
a^\dagger=\frac{1}{\sqrt2}\left(-\; \frac{d}{dx}+x\right),
\label{lad}
\end{equation}
relation (\ref{ccr}) takes the form of 
the bosonic oscillator, or Weyl algebra:
\begin{equation}
[a,a^\dagger]=1.
\label{weyl}
\end{equation}
The number operator $N\equiv a^\dagger a$  
satisfies the relations $[N,a^\dagger]=a^\dagger,\; [N,a]=-a$.
In appropriate units the 
Hamiltonian of a harmonic oscillator is equal to $N$ up to a 
constant term:
$2H=\{a^\dagger, a\}=-d^2/dx^2+x^2.$
The energy spectrum and orthonormal eigenstates are:
$$H|n\rangle=E_n|n\rangle,\qquad E_n=n+1/2, \qquad
|n\rangle=\frac{1}{\sqrt{n!}}(a^\dagger)^n|0\rangle,$$
where the vacuum state $|0\rangle$ is defined from the equation
$a|0\rangle=0,\; \langle x|0\rangle=\pi^{-1/4} \exp (-x^2/2)$ (we set
the phase of this state equal to zero).

Coherent states of the harmonic oscillator may be defined
either as eigenstates of the annihilation operator $a$:
\begin{equation}
a|\alpha\rangle=\alpha|\alpha\rangle,
\label{def1}
\end{equation}
or as a result of the application of the displacement operator
to the vacuum:
\begin{equation}
|\alpha\rangle=
e^{\alpha a^\dagger-\alpha^*a}|0\rangle=
e^{-|\alpha|^2/2} \sum_{n=0}^\infty \frac{\alpha^n}{\sqrt{n!}}|n\rangle.
\label{def2}
\end{equation}
Both definitions are essentially equivalent and give:
\begin{equation}
\psi_\alpha(x)\equiv 
\langle x|\alpha\rangle=\pi^{-1/4}\exp \left(\frac{\alpha^2-|\alpha|^2}{2}-
(\frac{x}{\sqrt2}-\alpha)^2\right).
\label{states}
\end{equation}
The states (\ref{def1}) are defined up to a phase factor
$\exp i\chi(\alpha, \alpha^*)$, where $\chi(\alpha, \alpha^*)$ is an 
arbitrary real function such that $\chi(0, 0)=0$. Only a special choice
of $\chi$ corresponds to (\ref{states}). Note that the shift of $x$
by a real constant $x_0$ is a canonical transformation which is not
completely equivalent to the shift of $\alpha$:
$$
\psi_\alpha(x-x_0)=e^{ix_0{\rm Im }\; \alpha/\sqrt2}\psi_{\alpha+x_0/\sqrt2}(x).
$$
Since the bound states wave functions are expressed through the Hermite
polynomials $H_n(x)$,
\begin{equation}
\langle x|n\rangle= {H_n(x)\over\sqrt{2^n n!\sqrt{\pi}}}e^{-x^2/2},
\label{herm}
\end{equation}
the relations (\ref{def2}), (\ref{states}) lead to the generating 
function for these polynomials
$$
\sum_{n=0}^\infty{t^n\over n!}H_n(x)=e^{2xt-t^2}, \qquad t={\alpha\over\sqrt2}.
$$

The strongest quantitative measure 
of differences in the behavior of quantum and classical
particles is expressed by the Schr\"odinger-Robertson 
uncertainty principle \cite{sudar}:
\begin{equation}
\Delta \equiv \sigma_{xx}\sigma_{pp}-\sigma_{xp}^2\geq \frac{1}{4},
\label{uncer}
\end{equation}
where $\sigma_{bc}=\frac{1}{2}\langle bc+cb\rangle-\langle b\rangle
\langle c\rangle,$
and the angular brackets denote averaging over an arbitrary normalizable 
state for which the mean values are well defined,
$\langle b\rangle=\langle\psi| b|\psi\rangle.$

Averaging over the coherent states $|\alpha\rangle$ one finds: 
$\langle x^2\rangle=\frac{1}{2} +\langle x\rangle^2,\;
\langle p^2\rangle=\frac{1}{2} +\langle p\rangle^2,\;
\langle px+xp\rangle=2\langle x\rangle\langle p\rangle,$
where
$$
\langle x\rangle=(\alpha+\alpha^*)/\sqrt2, \qquad
\langle p\rangle=(\alpha-\alpha^*)/i\sqrt2.
$$
In this case $\Delta=1/4$. 
Since $\sigma_{xp}=0$, this is the lower bound of the 
Heisenberg's uncertainty relation as well: $\sigma_{xx}\sigma_{pp}=1/4.$
Note that the latter equality does not determine uniquely coherent states. 
One may scale $x$ in (\ref{states}) by a real number and preserve 
minimality of the product $\sigma_{xx}\sigma_{pp}$. The resulting states
are called squeezed states; a more wide class of states corresponds 
to the lower bound of (\ref{uncer}). It is necessary to impose 
additional constraints in order to get (\ref{states}) uniquely 
\cite{sudar,nieto}.

The elementary example considered here illustrates three possible definitions
of coherent states for an arbitrary system: 1) as eigenfunctions
of some symmetry operators lowering the energy, 2) as an
orbit of states generated by a chosen group element from a fixed state,
and 3) as  minimum uncertainty states for some physically significant
operators. It is the first definition that we employ in this paper.

\section{Parity invariance and superpositions of coherent states}
\setcounter{equation} 0

The formulas (\ref{lad})-(\ref{states}) are well known and widely used in 
quantum physics. First, we would like to find their analogs for a
non-standard realization of the bosonic oscillator algebra. 
It is easy to see that this algebra has a nontrivial 
automorphism (i.e. a map onto itself), or canonical transformation
associated with the parity operator $P$:
\begin{equation}
PxP=-x,\qquad PpP=-p,\qquad P^2=1,\qquad P^\dagger=P.
\label{parity}
\end{equation}
Moreover, the transformation of $x$ and $ p$ to the hermitian variables 
$$
\tilde x\equiv -ipP,\qquad \tilde p\equiv ixP,
$$
is also canonical: $[\tilde x,\tilde p]=i.$
Although this is a quite simple fact, it leads to the non-trivial 
reshaping of coherent states.
Let us define new creation and annihilation operators:
\begin{equation}
A^\dagger=\frac{1}{\sqrt2}\left(-\; \frac{d}{dx}+x\right)P, \qquad
A=\frac{1}{\sqrt2}P\left(\frac{d}{dx}+x\right),
\label{newladder}
\end{equation}
so that $A+A^\dagger=\sqrt2\tilde x,\; A-A^\dagger=i\sqrt2 \tilde p.$
Evidently, the algebra, Hamiltonian, and vacuum state of the harmonic 
oscillator defined by the operator $A$ coincide
with the ones considered in the preceeding section.
However, there is an essential difference in the structure of
energy eigenstates generated by $A^\dagger$:
\begin{equation}
|n\rangle_{new}\equiv \frac{1}{\sqrt{n!}}(A^\dagger)^n|0\rangle
= (-1)^{s_n}|n\rangle,
\label{hermnew}
\end{equation}
where the sign factor $(-1)^{s_n}$ has the form:
\begin{equation}
(-1)^{s_n}=\cases{1, \qquad\quad  n=4k,\; 4k+1,\qquad\qquad
k=0,1,2,\dots\cr
-1, \qquad \; n=4k+2,\; 4k+3.}
\label{sign}
\end{equation}
This follows from the relation $(A^\dagger)^2=-(a^\dagger)^2$
and the parity invariance of the vacuum.

Denote by $|\alpha\rangle_{P}$ the eigenstates of $A$, 
$A|\alpha\rangle_{P}=\alpha|\alpha\rangle_{P},$ or
\begin{equation}
(d/dx+x)\psi_{\alpha}^P(x)= \sqrt{2}\alpha\psi_{\alpha}^P(-x),
\qquad \psi_{\alpha}^P(x)=\langle x|\alpha\rangle_P,
\label{diffdiffeq}\end{equation}
This is not an ordinary differential equation, but it can be easily solved
using the relation $A^2=-a^2$. Picking out the appropriate combination 
of two linearly independent eigenstates of $a^2$ with the eigenvalue
$-\alpha^2$, we find:
\begin{equation}
|\alpha\rangle_P=\frac{1}{\sqrt2}\left(e^{-i\pi/4}|i\alpha\rangle
+e^{i\pi/4}|-i\alpha\rangle\right),
\label{newcoh}
\end{equation}
$$_P\langle\beta|\alpha\rangle_P=\langle\beta|\alpha\rangle=
\exp (\beta^*\alpha-\frac{1}{2}|\alpha|^2-\frac{1}{2}|\beta|^2),$$
where $|\alpha\rangle$ are the canonical coherent states. 
In the coordinate representation one has
\begin{equation}
\psi^P_\alpha(x) = 
{\sqrt2\over \pi^{1/4}} \exp\left({\alpha^2-|\alpha|^2-x^2\over 2}\right)
\cos(\sqrt2 \alpha x - {\pi\over 4}).
\label{newcohcoord}
\end{equation}
We call (\ref{newcohcoord}) the {\it parity coherent states}, 
because the parity symmetry plays a central role in their definition.
As it is argued below, analogous states can be constructed for 
an arbitrary symmetric potential.
Since $|\alpha\rangle_P$ is defined by (\ref{def2}) with $A,\; A^\dagger$
instead of $a, \; a^\dagger$, one can derive the following generating relation
for the Hermite polynomials
$$
\sum_{n=0}^\infty {(-1)^{s_n}t^n\over n!} H_n(x)=
\sqrt2 e^{t^2}\cos\left(2tx - {\pi\over 4}\right).
$$

The $|\alpha\rangle_P$'s are not minimal uncertainty states for
the variables $x$ and $p$ when $\alpha\neq 0$:
$$\langle x\rangle=\frac{\alpha+\alpha^*}{\sqrt2}e^{-2|\alpha|^2},
\qquad\langle p\rangle=\frac{\alpha-\alpha^*}{i\sqrt2}e^{-2|\alpha|^2},$$
\begin{equation}
\sigma_{xx}=
(1-(\alpha-\alpha^*)^2-(\alpha+\alpha^*)^2e^{-4|\alpha|^2})/2,
\label{sigxx}
\end{equation}
$$
\sigma_{pp}=
(1+(\alpha+\alpha^*)^2+(\alpha-\alpha^*)^2e^{-4|\alpha|^2})/2,
$$
$$
\sigma_{xp}=({\alpha^*}^2-\alpha^2)(1+e^{-4|\alpha|^2})/2i,
$$
so that
$$
\Delta=\frac{1}{4}(1+\rho(1-(1+\rho)e^{-\rho})),\qquad \rho\equiv 4|\alpha|^2,
$$
and the minimum is reached only for $\rho=0$. 
However by construction itself, the parity coherent states
minimize the product of uncertainties in the new canonical variables 
$\tilde x$ and $\tilde p$. 

The minimum of $\sigma_{xx}$ (\ref{sigxx}) is reached for 
$\alpha=\alpha^*=1/2$: $\sigma_{xx}\big|_{min}=(1-e^{-1})/2\approx 0.32,$
i.e. there is a squeezing for small $|\alpha|$. However, the states 
$|\alpha\rangle_P$ differ from the squeezed states \cite{nieto} arising as 
eigenstates of the annihilation operator after the canonical transformation 
$$
A=S^\dagger a S=\cosh |z|\; a+{z\over |z|}\sinh |z|\; a^\dagger, 
$$
generated by the unitary operator 
\begin{equation}
S(z)=e^{(za^{\dagger 2}-z^*a^2)/2}.
\label{squeeze}
\end{equation}
If one multiplies this $A$ by the parity operator from the left,
then eigenfunctions of the resulting operator 
will be given again by superposition (\ref{newcoh}), 
but now with the $|\alpha\rangle$'s on the r.h.s. being replaced
by squeezed states. Note that in the Bargmann-Fock representation
of the harmonic oscillator algebra, when $A=Pd/dz,\; A^\dagger=zP$, 
one has $\psi_\alpha^P(z)\propto \cos(\alpha z-\pi/4)$.

It is easy to construct analogs of (\ref{newcoh}) for the 
two oscillator algebra: $[a_j,a_k^\dagger]=\delta_{jk},\;
[a_j,a_k]=0, \; j,k=1,2.$
Again, the combinations $A_j=Pa_j$, where $P$ is an operator which
inverts both space coordinate axes, satisfy the same algebra. Such
a transformation affects only the sign of the energy eigenfunctions.
Eigenstates of the $A_j$-operators, 
$A_j|\alpha_1,\alpha_2\rangle_P = \alpha_j|\alpha_1,\alpha_2\rangle_P,$ 
are given by the superposition:
\begin{equation}
|\alpha_1,\alpha_2\rangle_P=
\frac{1}{\sqrt2}\left(e^{-i\pi/4}|i\alpha_1\rangle_1|i\alpha_2\rangle_2+
e^{i\pi/4}|-i\alpha_1\rangle_1|-i\alpha_2\rangle_2\right),
\label{ecs}
\end{equation}
where $|\alpha\rangle_j$ are the canonical coherent states of the 
$j$-th degree of freedom. This formula is obtained by choosing
an appropriate linear combinations of the $a_j^2$-operator eigenstates.
Generalization of (\ref{ecs}) to an arbitrary number of oscillators 
is obvious.

The unitary operator $U$ that transforms (\ref{lad}) into (\ref{newladder}) 
(or its multidimensional analog) is easily
found due to the relation between the operators $P$ and $N$
in the Hilbert space:
$P=(-1)^N=\exp i\pi N$. It is 
$$
U=e^{\pi iN(N-1)/2}, \qquad A=U^\dagger a U, \quad
A^\dagger = U^\dagger a^\dagger U.
$$ 
This is a special case
of the canonical transformations generated by the operator 
\begin{equation}
U=e^{-i\theta(N)}, \qquad U^\dagger aU=e^{i\varphi(N)}a, \qquad
\varphi(N)=\theta(N)-\theta(N+1),
\label{unitop}
\end{equation}
where $\theta(N)$ is a non-singular function
(such transformations were discussed in \cite{fivel,brandt} without
relating them to superpositions of coherent states).
In the classical case, when $H_{cl}=\alpha^*\alpha, \;
\alpha=(x+ip)/\sqrt2$, relations (\ref{unitop}) correspond to 
a rotation of $\alpha$ through an angle depending on energy:
$U^\dagger \alpha U=e^{i\varphi(\alpha^*\alpha)}\alpha$ which is
obviously a symmetry of the system. 

Applying the operator $U^\dagger$ (\ref{unitop}) to $|\alpha\rangle$, 
one finds eigenstates of the operator $A=e^{i\varphi(N)}a$:
\begin{equation}
|coh\rangle=e^{-|\alpha|^2/2}\sum_{n=0}^\infty \frac{\alpha^n}{\sqrt{n!}}
e^{i\theta(n)}|n\rangle.
\label{tit}
\end{equation}
These are the generalized coherent states of Titulaer and Glauber 
\cite{titulaer}. The states (\ref{newcoh}) belong to their particular
subclass characterized by the periodicity 
condition 
\begin{equation}
\theta(n+M)=\theta(n),
\label{period}
\end{equation}
imposed upon
the function $\theta(n)$ (in our example $M=4$, cf. (\ref{sign})). 
In this case, one has \cite{birula,stoler}
$$
|coh\rangle=\sum_{k=0}^{M-1} C_k|\epsilon^k\alpha\rangle,
\qquad \epsilon=e^{2\pi i/M}, \quad \epsilon^M=1,
$$  
i.e. $|coh\rangle$ is a superposition of $M$ coherent states with the
parametrizing variable $\alpha$ modulated by the powers of a
primitive $M${th} 
root of unity. The states $|\epsilon^k\alpha\rangle$ are linearly independent,
so that they can be orthonormalized. For the simplest $M=2$ case
these orthonormal superpositions take the form of even and odd coherent
states \cite{dodonov}
$$
|\alpha_\pm\rangle=
{|\alpha\rangle\pm |-\alpha\rangle \over
\sqrt{2\pm 2e^{-2|\alpha|^2 }}}.
$$
The general root-of-unity analogs of these states $|\alpha_l\rangle,$
$l=0, 1, \dots, M-1$, 
$\langle\alpha_l|\alpha_m\rangle=\delta_{lm},$ have the following form
\begin{eqnarray}
|\alpha_l\rangle &=& C_l(\alpha)\sum_{m=0}^{M-1}\epsilon^{-lm}
|\epsilon^m \alpha\rangle
\label{rootstates}\\
&=& M C_l(\alpha)e^{-|\alpha|^2/2}\sum_{k=0}^\infty
{\alpha^{Mk+l}\over\sqrt{(Mk+l)!}}|Mk+l\rangle, 
\nonumber \\
|C_l(\alpha)|^2 &=&{e^{|\alpha|^2}\over M}\left( 
\sum_{m=0}^{M-1}\epsilon^{-lm} \exp(\epsilon^m|\alpha|^2)\right)^{-1}.
\nonumber 
\end{eqnarray}
It is easy to derive these formulas using the 
kernel of the finite-dimensional Fourier transformation $F$,
$$
(F)_{lm}\equiv {1\over\sqrt{M}}\epsilon^{lm}, \qquad F^\dagger F=F F^\dagger =1,
$$
$$
F^t=F, \qquad 
(F^2)_{lm}=\delta_{l0}\delta_{m0}+\delta_{M-l, m}, \qquad F^4=1.
$$

The states  (\ref{rootstates}) do not belong to the class (\ref{tit})
because  for arbitrary values of $\alpha$ they cover only a part of the
Hilbert space,
$$
\int d^2\alpha\; |\alpha_l\rangle\langle\alpha_l|=\pi_l, \qquad
d^2\alpha\equiv d(Re\;\alpha)d(Im\;\alpha)/\pi,
$$
$$
\pi_l\pi_m=\pi_m\delta_{lm}, \qquad \sum_{l=0}^{M-1} \pi_l=1,
$$
whereas the Titulaer-Glauber coherent states form an overcomplete 
set of states.
The possibility to split Hilbert space of the harmonic oscillator 
to an arbitrary number of 
orthogonal subspaces is related to the fact that the projection operators
$\pi_l$ are conserved, 
$$
\pi_l={1\over M}\sum_{m=0}^{M-1}\epsilon^{m(N-l)}, \qquad [H, \pi_l]=0.
$$

The distinguished property 
of the parity operator is that it is the only symmetry of separate
kinetic and potential terms of the harmonic oscillator Hamiltonian.
Moreover, the definition $Pf(x)=f(-x)$ works for any function
independently of its membership in the Hilbert space
(for complex $x$ this is just the rotation of the complex plane by $\pi$). 
It is only for the functions which can be expanded over the normalizable states
$|n\rangle$ that the operator $\exp i\pi N$ coincides with parity, 
e.g. for the  non-normalizable
eigenfunctions of Hamiltonian the action of this operator is formally 
equivalent to multiplication by some phase factor which is not related to the 
transformation $x\to - x$. Therefore it is not clear what are the analogs
of the operators $\exp{i\theta(N)}$ beyond the Hilbert space context.

Although the finite-dimensional truncation of the Titulaer-Glauber
coherent states was discovered a long time ago \cite{birula,stoler}, 
only recently in \cite{yurke} have Yurke and Stoler derived explicitly the
superposition (\ref{newcoh}), with $i\alpha$ being replaced by $\alpha$.
In their approach it emerged as a result of a fixed period
time evolution of the standard coherent states governed by a 
specific Hamiltonian $\propto N^k$, $k$ even. The squeezing and 
other properties of these and more general finite-term superpositions 
of coherent states have been analyzed in \cite{mec}-\cite{wang}.
The states (\ref{ecs}) belong to the class of so-called two-particle 
entangled coherent states \cite{mec,sanders,green}, 
which have the intrinsic property of non-factorizability
into the product of one-particle states. For the construction of
multimode analogs of the
even and odd coherent states, see \cite{ansari}. Recent interest in 
superpositions of macroscopically distinguishable quantum states
(Schr\"odinger cat states)
like (\ref{newcoh}) or the entangled states like (\ref{ecs})
is inspired by new possibilities to create them with the help
of optical techniques. This in turn provides interesting 
experimental tests of the basic principles of quantum mechanics
(see, e.g. \cite{yurke,mec,sanders,green,gravi} and references therein).

The general set of coherent states constructed with the help of canonical 
transformations based upon the parity operator has the following form.
Consider  the unitary operator $V$,
\begin{equation}
V=\cos\varphi +iP\sin\varphi, \qquad V^\dagger V=VV^\dagger=1,
\label{generalparityop}
\end{equation}
where $\varphi$ is an arbitrary parameter, and construct the 
Weyl algebra generators analogous to (\ref{newladder}):
\begin{equation}
A=Va=aV^\dagger, \qquad A^\dagger=a^\dagger V^\dagger=Va^\dagger,
\label{paritygen}\end{equation}
$$
[A, A^\dagger]=[a,a^\dagger]=1,\qquad A^2=a^2,\qquad 
(A^\dagger)^2 =(a^\dagger)^2.
$$
The eigenfunctions of the operator $A$, or parity coherent
states, are easily found 
\begin{equation}
\psi_\alpha^P(x)=\frac{1}{2}\left((1+e^{-i\varphi})\psi_\alpha(x)+
(1-e^{-i\varphi})\psi_{-\alpha}(x)\right),
\label{generalparity}
\end{equation}
with $\psi_\alpha(x)$ defined in (\ref{states}).
Calculating the uncertainties of $x$ and $p$ in these states,
we obtain
$$
\Delta(\varphi)=\frac{1}{4}\left(1 + 
\rho\sin^2\varphi(1-(1+\rho)e^{-\rho})\right),
\qquad \rho=4|\alpha|^2.
$$
It is seen that the choice $\varphi=\pi/2$, which corresponds to the
Yurke-Stoler coherent states,   is extremal -- for it the
value of $\Delta$ is maximally deviated from the standard
coherent states case $\varphi=0$, when $\Delta=1/4$. 

Let us discuss the Titulaer-Glauber states when the
phases $\theta(n)$ satisfy the following $q$-periodicity condition:
\begin{equation}
\theta(n+M)=q\theta(n),
\label{qperiod}
\end{equation}
which is a simple $q$-deformation of the condition (\ref{period}).
When $q\to1$ one can renormalize $\theta$, $\theta=\tilde\theta-\phi/(1-q)$,
and get $\tilde\theta(n+M)=\tilde\theta(n)+\phi$, which is a quasiperiodicity 
condition. However, the effect of such a shift by $\phi$ is equivalent
to the multiplication of $\alpha$ by the factor $\exp{i\phi/M}$ which
is harmless for the representation of $|coh\rangle$ as a 
finite-term superposition of canonical coherent states.

For generic values of $q$ there is no split of $|\alpha\rangle$ onto 
a superposition of a finite number of coherent states. For 
$M=1$ one has $\theta(n)=\phi q^n$, and
\begin{eqnarray}
|coh\rangle &=&e^{-|\alpha|^2/2}\sum_{n=0}^\infty {\alpha^n\over \sqrt{n!}}
e^{i\phi q^n}|n\rangle  
= \sum_{k=0}^\infty {(i\phi)^k\over k!}|q^k\alpha\rangle.
\label{qperiodcs}
\end{eqnarray}
However, when the parameter $q$ is a primitive $M${th} root of unity,
$q^M=1$, the sum (\ref{qperiodcs}) is truncated:
$$
|coh\rangle=\sum_{l=0}^{M-1} B_l(\phi) |q^l\alpha\rangle,
\qquad B_l(\phi)={1\over M}\sum_{m=0}^{M-1}q^{-lm}e^{iq^m\phi}.
$$
The coefficients $B_l$ are defined by the finite sums of exponentials
which were already encountered in the calculation of
normalization constants for the orthonormal states $|\alpha_l\rangle$
(\ref{rootstates}). 
We would like to note that the $q$-deformation (\ref{qperiod})
is not related to the $q$-coherent states to be discussed below,
although the discrete energy
spectrum of the corresponding Hamiltonians is found from the similar
formula $E_{n+M}=q^2E_n$.

\section{Universality of superpositions of coherent states}
\setcounter{equation} 0

Let us discuss whether the superpositions of a 
finite number of coherent states considered above 
are characteristic only for the harmonic oscillator case or their
form carries universal character applicable to any system.
Unfortunately, there is no completely satisfactory definition of 
coherent states $|\alpha\rangle$ for an arbitrary Hamiltonian
\begin{equation}
H=-\;{d^2\over dx^2} + u(x),
\label{hamgen}
\end{equation}
even if the potential $u(x)$ is an analytical or infinitely differentiable 
function of $x$ (i.e. when the random, singular, and potentials from 
$C^k,\; k<\infty$, are excluded). Denote by $|\lambda\rangle$ the physical 
eigenfunctions of the Hamiltonian:
$$
H|\lambda\rangle=E(\lambda)|\lambda\rangle, 
$$
where $\lambda$ is some index labeling the spectrum; 
$\lambda\equiv n=0, 1, \dots$ for the discrete eigenvalues
(the ordering $E(n) < E(n+1)$ is assumed), and for the continuous
spectrum $\lambda$ may be thought of as some continuous variable 
such that the spectrum $E(\lambda)$ is monotonically covered by variation of
$\lambda$, say from $\lambda_0$ to $\infty$.
We assume that the continuous spectrum states are normalized by the condition
$\langle\lambda|\sigma\rangle\propto \delta(\lambda-\sigma).$
From the completeness of generalized eigenfunctions of $H$ in Hilbert space,
\begin{equation}
\sum_{n=0}^{n_b - 1} |n\rangle\langle n|+\int_{\lambda_0}^\infty d\lambda
|\lambda\rangle\langle\lambda| = 1,
\label{completeness}\end{equation}
where $n_b$ is the number of bound states, 
it follows that for any definition of coherent states $|\alpha\rangle$, 
these states can be expanded over $|n\rangle$ and $|\lambda\rangle$:
\begin{equation}
|\alpha\rangle=\sum_{n=0}^{n_b-1} c_n(\alpha)\;|n\rangle
+\int_{\lambda_0}^\infty  d\lambda \; c_\lambda(\alpha)\;|\lambda\rangle.
\label{cohgen}
\end{equation}
The continuous spectrum exists whenever, for $x\to\infty$ (or $-\infty$),
the potential is not bounded from below or it is bounded from above.
In these cases one cannot exclude in general the
second term in expansion (\ref{cohgen}).
Such possibilities are rarely discussed in the literature;
some models of coherent states built only from the 
continuous spectrum states are described in the next three sections. 

Suppose for a moment that the Hamiltonian $H$ has only a discrete spectrum,
i.e. $n_b=\lambda_0=\infty$. Then, following many existing examples \cite{coh},
it is natural to assume that coherent states $|\alpha\rangle$ represent a
generating function for stationary energy states of the form
\begin{equation}
|\alpha\rangle=\sum_{n=0}^{\infty}\alpha^n c_n(|\alpha|) |n\rangle.
\label{nexpansion}
\end{equation}
Since the coefficients $c_n$ depend only on the modulus of $\alpha$,
these states are complete,
$$
\int {d^2\alpha}\rho(|\alpha|) |\alpha\rangle\langle\alpha|=
\sum_{n,m=0}^{\infty} |n\rangle\langle m|\int d^2\alpha 
\rho(|\alpha|)\alpha^{* m}\alpha^n c_nc_m^*=1,
$$
provided there exists a measure density $\rho(|\alpha|)$ which satisfies
the relations
$$
\int_0^\infty dr \rho(r)r^{2n+1} |c_n(r)|^2={1\over 2}
$$
for arbitrary $n=0, 1, \dots$. When the variables of $c_n(r)$ separate,
$c_n(r)=c_ng(r)$, which is typical for the ladder operator approach,
this is a moment problem.
Theoretically it is possible that $\rho(r)$
is defined non-uniquely in which case there are many physically distinguishable
representations of observables in the coherent states basis.

An interesting  definition of coherent states
based on the uncertainty principle was suggested in \cite{nietosim}
for a wide class of potentials.
It uses the fact that for a classical particle moving in a
convex potential one can make a non-canonical non-linear change of 
phase space variables such that in the new ``coordinates" the 
particle's dymanics is described by the harmonic oscillator equations of 
motion. After quantization, the minimum uncertainty states of these
harmonic motion ``position" and ``momentum" operators are called 
coherent states.
By definition, this procedure is tied to the quasiclassical approximation.
Its general group-theoretical meaning is not clear to the 
author, probably the approximate dynamical symmetry approach of \cite{fivel}
can be useful in this context.  Another constructive definition of coherent 
states for generic discrete spectrum systems has been
suggested by Klauder \cite{klaudernew}. In this approach, the requirement that
time evolution of coherent states be equivalent to the change 
$\alpha\to e^{i\omega t}\alpha$ is taken as a basic property.
Such states do not spread but for them one has more complicated 
expansions than (\ref{nexpansion}).

In the following we utilize a different definition of 
coherent states; namely, we assume that they 
are defined as eigenstates of some lowering operator $A$,
i.e. the operator which maps a part of physical solutions of a given stationary 
Schr\"odinger equation to the physical ones with lower energy.
Even for systems with purely discrete spectrum this requirement does 
not necessarily mean that 
$A$ annihilates the ground state or that $A$ is 
the lowering operator mapping discrete spectrum eigenfunction $|n\rangle$ to 
the closest from below state $|n-1\rangle$ (it may jump over some physical
states which is always so for the continuous spectrum). For example, 
similar to the situation in \cite{s1,s2}, $A$ may play the role of both 
lowering and raising  operators for different ranges of energy.

We use the factorization method \cite{infeld,miller} as a tool for searching 
for such a symmetry operator $A$. Let us factorize the Hamiltonian 
(\ref{hamgen}), i.e.
represent it as a product of two first order 
differential operators conjugated formally to each other: 
\begin{equation}
H=a^\dagger a+E_0, \qquad 
a=d/dx+f(x),\quad a^\dagger=-d/dx+f(x),
\label{factorization}\end{equation}
where $E_0$ is some constant. 
The potential $u(x)$ and superpotential $f(x)$ are related 
by the Riccati equation, $u(x)=f^2(x)-f'(x)+E_0$.   
The zero mode of $a$, 
$$
a\psi_0(x)=0, \qquad 
\psi_0(x) \propto e^{-\int^x f(y)dy},
$$
is the generalized eigenfunction of the Hamiltonian with 
the eigenvalue $E_0$. When this function is normalizable and nodeless,
$E_0$ is the ground state energy. 
If $a$ is a symmetry operator, i.e. if it maps physical eigenstates 
of the Hamiltonian $H$ onto themselves, 
then coherent states could be defined as eigenstates of $a$, 
but there may be non-uniqueness even in this simple
situation. Indeed, the factorization involves an arbitrary unitary 
operator $T$:
\begin{equation}
A^\dagger A=a^\dagger a, \qquad
A^\dagger=a^\dagger T, \quad A=T^{-1}a,\quad T^\dagger=T^{-1}.
\label{freedom}\end{equation}
Actually, one may write $A^\dagger=a^\dagger C$,
$A=Da,$ where $C$ and $D$ are two operators satisfying $CD=1$. The operators
$A^\dagger$ and $A$ are conjugated to each other if $C=D^\dagger$, i.e.
when $D$ is an isometric operator \cite{pursey}. Only the additional 
requirement $D D^\dagger =1$ makes $D$ unitary, we restrict ourselves to 
this case.

Coherent states of potentials for which 
$A=T^{-1}a$ is a symmetry operator, for some unitary operator $T$, 
are thus defined by the equation: 
\begin{equation}
\psi_0(x) \frac{d}{dx} \frac{\psi_\alpha(x)}{\psi_0(x)}=
\alpha T\psi_\alpha(x),
\label{general}
\end{equation}
where $\psi_0(x) $ is an eigenstate of the Hamiltonian (not necessarily
a physical one). 
The natural extension of this definition involves on the l.h.s. 
of (\ref{general}) a differential operator of the $N${th} order. 
In that case $A^\dagger A$ is equal to an order $N$ polynomial of 
a Hamiltonian, i.e. one has a generalized factorization scheme 
which will be described in section 6. 

Suppose now that the expansion (\ref{nexpansion}) takes place
(it does not mean that there is no continuous spectrum, there may be
accumulation points such that the set $|n\rangle$ is closed under
the action of $A$). Then, one can introduce the
formal number operator, $N$, satisfying $N|n\rangle=n|n\rangle$. 
Acting upon such $|\alpha\rangle$ by the unitary operator 
$U^\dagger=e^{i\theta(N)}$,
one gets the Titulaer-Glauber type coherent states 
for the chosen class of potentials.
The periodicity condition $\theta(n+M)=\theta(n)$ leads again to
the finite term superpositions of $|\epsilon^k\alpha\rangle$.
Therefore the form (but not the normalization constants)
of the superpositions $|\alpha_k\rangle$
(\ref{rootstates}) is universal -- for any $H$ they perform a split
of discrete spectrum Hilbert subspace onto orthogonal components. 
Note however, that the
symmetry properties of these superpositions may be different; in particular,
the general even and odd coherent states are not eigenstates of the 
parity operator. For instance, for the shifted harmonic oscillator potential 
$u(x)=(x-x_0)^2$ one has:
\begin{equation}
\langle x|\alpha_\pm\rangle\propto 
e^{-(x-x_0)^2/2}\left( e^{\sqrt2 (x-x_0)\alpha}\pm e^{-\sqrt2 (x-x_0)\alpha}
\right),
\label{shift}\end{equation}
which are eigenstates of the hermitian conserved charge $P\exp (2x_0d/dx)$. 
For the general asymmetric potential, 
$|\alpha_\pm\rangle$ are eigenstates of the abstract operator $\exp i\pi N$,
which does not have a simple form in the coordinate representation.

Similarly, one can always define upon the discrete spectrum 
an analog of the Yurke-Stoler coherent states, but they will be eigenstates of
the operator $Pa$, an analog of (\ref{newladder}), only in the case 
when $Pa=- aP$, which assumes that the potentials are symmetric, 
$u(-x)=u(x)$ (if one has $Pa=aP$, then the symmetric and antisymmetric
eigenfunctions of $a$ diagonalize simultaneously the operator $Pa$). 
We come thus to the conclusion that the notion of ``parity coherent states"
is less universal than that of even and odd coherent states and their
higher root-of-unity generalizations defined by the abstract formula
(\ref{rootstates}). The special name for the states (\ref{generalparity}), 
and similar ones, was introduced in order to
distinguish Titulaer-Glauber states
with the $M=2$ periodic phases for symmetric and asymmetric
potentials (in the first case there is no need for the abstract operator $N$). 

Single-valuedness of the expansion (\ref{nexpansion}) under the rotation
$\alpha\to e^{2\pi i}\alpha$ is the key property allowing to build finite-term
superpositions of coherent states (\ref{rootstates}). In general such 
property does not hold. In the next section we consider a model 
where only the continuous spectrum piece is present in (\ref{cohgen})
with $c_\lambda(\alpha)\propto \alpha^{-\ln\lambda/\ln q^2}$. In this case 
the change $\alpha\to \epsilon^k \alpha$, $\epsilon^M=1$, does not provide 
a split of Hilbert space onto the finite number of orthogonal components.

\section{Coherent states of the free particle}
\setcounter{equation} 0

The above definition of coherent states (\ref{general})
is applicable to systems with continuous energy spectra. 
Consider the simplest possible case of zero potential, $H=-d^2/dx^2$, 
for which the generalized eigenfunction of lowest energy is a simple constant. 
Solutions of the 
Heisenberg equations of motion $x=p_0t+x_0,\; p=p_0$, where $t$
is the time variable and $p_0, x_0$ are operators at $t=0$, 
are identical with the classical ones.
If this coincidence would be taken as the basis for the definition of coherent
states, then any normalizable state of the free particle has to be 
considered as coherent. However despite the same form for the equations 
of motion, even for the $\langle p_0\rangle=0$ case, when the
classical particle stays at the point $x_0$, the quantum 
particle tries to occupy the whole space: 
$\sigma_{xx}\propto \langle p^2_0\rangle t^2$, $t\to\infty$, 
and non-spreading wave packets do not exist. 

The simplest factorization of the free particle Hamiltonian is obvious and we 
have $a=d/dx$. Let us find eigenfunctions of $A=T^\dagger d/dx$ when $T$ 
is a unitary operator performing an affine transformation,
or the parity operator.
If $T=1$, then $\psi_\alpha(x)\propto e^{\alpha x}$ which
are bounded (but unnormalizable) functions for purely imaginary 
$\alpha$: $\alpha=ip,\; -\infty < p < \infty,$ so that
$\psi_{ip}(x)$ coincide with the momentum eigenstates. 
If $T$ is a translation by $h$ operator, then 
$$
d \psi_\alpha(x)/dx=\alpha\psi_\alpha(x+h)
$$ 
and again there are simple solutions of the form 
$\psi_\alpha(x)\propto e^{ip x}$ but now $\alpha$ has a real part:
$\alpha=ip e^{-iph}$. If $T=P$, the parity operator, then 
one gets superposition (\ref{newcoh}), 
$\psi_\alpha^P(x)\propto \cos(\alpha x-\pi/4)$, $\alpha$ real.
Since these states are not physically realizable (they belong to the
continuous spectrum), their ``coherence" is formal. This is related
to the fact that the chosen symmetry operators $A$
are integrals of motion commuting with the Hamiltonian. 
As is seen from the considerations given below, such $A$'s appear from 
real lowering operators in a special limit such that, in fact, all
Hamiltonian eigenstates acquire 
a flavor of coherent states (this is reminiscent to the 
approach of \cite{malkin}, where coherent states are defined 
as eigenstates of some combinations of integrals of motion).
Examples of symmetry algebras for which the Hamiltonian eigenstates 
might be counted among the coherent states are given below. 
In the corresponding cases the operator $A$ may be simultaneously the 
lowering and raising operator, and 
an integral of motion for different parts of the spectrum.

Let $T$ be the scaling operator:
\begin{equation}
Tf(x)=\sqrt{|q|}f(qx), 
\label{scaling}
\end{equation}
where $q$ is some real parameter, $0<q^2<1$. 
For positive $q$, $T$ is just the squeezing operator in a special form:
$T=S(z=\ln q)=q^{(a^{\dagger 2} - a^2)/2}$, where $a^\dagger, a$
are given by (\ref{lad}); and for negative $q$ it is a
product of the same $S$ and the parity operator: $T=SP$.
For simplicity we assume in this section that $q$ is positive. 
Now the operator $A=T^\dagger a$ is not an integral of motion,
but the {\it raising} operator for the $\lambda > 0$ solutions of the 
Schr\"odinger equation
\begin{equation}
H\psi(x)=-\psi''(x)=\lambda \psi(x).
\label{freeeq}\end{equation}
The commutation relations
$$
A A^\dagger=q^2 A^\dagger A,\qquad AH=q^2HA, \qquad HA^\dagger=q^2 A^\dagger H
$$
look similar to some of the defining relations of quantum groups 
\cite{drinfeld,jimbo}. In fact, they are the progenitors of $q$-deformed 
oscillator algebra (see below).
Eigenfunctions of the operator $A$ are determined by the differential equation
with deviating argument   
\begin{equation}
\psi'_\alpha(x)=\alpha\sqrt{q}\psi_\alpha(qx), \qquad 0<q<1.
\label{katoeq}\end{equation}
The initial value problem for (\ref{katoeq}) is qualitatively different
from that for the ordinary differential equations because the initial 
conditions have to be fixed now on the interval $[qx_0, x_0]$.
When $x_0$ is a fixed point of the scaling transformation, i.e.
$x_0= 0$ or $\infty$, this interval shrinks to one point.
When $x_0 = \infty$, one fixes solutions by taking the asymptotic
form of $\psi_\alpha(x)$ from some class of permitted functions \cite{kato}. 
For $x_0=0$ it is natural to impose the initial condition 
$\psi_\alpha(0)=\gamma < \infty$. Then equation (\ref{katoeq}) 
has the unique analytical solution:
\begin{equation}
\psi_\alpha(x)=
\gamma \sum_{n=0}^\infty \frac{q^{n(n-1)/2}}{n!} (\alpha\sqrt{q}x)^n,
\label{series}
\end{equation}
which is an entire function of $x$ for any finite $|\alpha|$.
However as is shown in \cite{kato} (see also \cite{iserles}), 
the $x\to\infty$ asymptotics of any 
solution of (\ref{katoeq}) is dominated by the factor $\exp (-\ln^2 x/\ln q^2)$,
i.e. all solutions grow at infinity so that the functions $\psi_\alpha(x)$
do not describe physical states. Actually, this could be expected from the
fact that we were diagonalizing the raising operator. 

Consider eigenfunctions of the lowering operator $A^\dagger$,
$$
A^\dagger\psi_\alpha(x)= \alpha \psi_\alpha(x),\qquad 
A^\dagger=-\;{d\over dx} T,
$$
or
\begin{equation}
\psi_\alpha'(x)=-\;\alpha q^{-3/2} \psi_\alpha(q^{-1}x), \qquad 0<q<1.
\label{cohfree}
\end{equation}
Formal series solution of this equation with the boundary condition
$\psi(0)=\gamma$ looks similar to (\ref{series}) with $q$ being replaced by 
$q^{-1}$. But the radius of convergence of this series is equal to zero,
i.e. there are no solutions analytical at zero. This does not mean that
there are no solutions at all, in \cite{kato} it was shown that in fact
there are infinitely many non-analytical solutions of (\ref{cohfree})
from $C^\infty$ 
satisfying boundary condition $\psi(0)=\gamma$. Moreover, for any $\alpha$
there are solutions with the $|x|\to\infty$ asymptotics 
$\propto \exp (\ln^2 |x|/\ln q^2),$ i.e there are  
functions $\psi_\alpha(x)$ which {\it are normalizable}. One can expand 
such $\psi_\alpha(x)$ over the basis of Hamiltonian eigenfunctions 
\begin{equation}
\psi_\alpha(x)=\int_0^\infty dp\; e^{ipx}\phi_\alpha(p),
\label{freecs}
\end{equation}
which is the positive momentum part of the standard Fourier integral. 
Similarly one can consider the expansion over $e^{-ipx}$
(for negative $q$ the two regions of $p$ should be considered simultaneously).
Substituting this expression into (\ref{cohfree}) and solving the
corresponding finite-difference equation for the formfactor $\phi_\alpha(p)$, 
we find
\begin{equation}
\psi_\alpha(x)=\int_0^\infty dp\; e^{ipx} h(p)\exp 
{\ln^2 p/i\alpha \over 2\ln q}, 
\label{freecsform}
\end{equation}
where $h(p)$ is an arbitrary function periodic on the logarithmic scale,
$h(qp)=h(p)$, normalized by the condition 
$$
\int_{-\infty}^\infty|\psi_\alpha(x)|^2dx=2\pi\int_0^\infty dp |h(p)|^2
\exp{\ln^2 p/|\alpha| - (\arg i\alpha)^2\over \ln q} = 1.
$$
Due to the freedom in choice of $h(p)$ these
states can take various forms. 

Let us take, for example, the following $h(p)$,
$$
h(p)=a\sum_{n=-\infty}^\infty q^n\delta(p-bq^n), 
$$
which does not correspond to normalizable $\psi_\alpha(x)$. 
Despite the formality of the consideration in this case, we obtain the 
free particle's ``coherent states" in the form of Dirichlet series 
(cf. \cite{derfel,iserles})
\begin{equation}
\psi_\alpha(x)=a\exp{\ln^2 b/i\alpha\over 2\ln q}
\sum_{n=-\infty}^\infty q^{n(n+2)/2} 
\left(b\over i\alpha\right)^n e^{ibq^n x}, 
\label{cohfreeser}\end{equation}
which defines a function bounded for any real $x$ and $0<|\alpha|<\infty$.
This is an example of the bounded but non-normalizable solution of 
(\ref{cohfree}). The derived expansion could be interpreted as a 
$q$-Fourier series for $\psi_\alpha(x)$ because the sum goes
over the trigonometric functions whose argument is modulated by the 
powers of $q$ (after renormalizations, $\sum c_n\exp(i\tilde x q^n)$ 
may become the standard Fourier series in the limit $q\to 1$ due to the 
relation $(q^n-1)/(q-1)\to n$, in our case this gives a divergent series).
This situation does not seem to be related to the $q$-Fourier 
transformation considered in \cite{askey}; it resembles more the 
wavelet transform \cite{meyer} or the generalized Taylor expansions
for atomic functions \cite{rvachev}.

In the above factorization of the free particle
Hamiltonian we took $E_0=0$. A much more rich situation arises when 
$E_0=-\beta^2$, where $\beta$ is a nonzero real number.
This gives $H=a^\dagger a -\beta^2,$ $a=d/dx+\beta$. Evidently $a$ is 
an integral of motion, but from the properties of affine 
transformation operator $T$,
\begin{equation}
Tf(x)=\sqrt{q}f(qx+l),\qquad 0<q< 1, \qquad -\infty<l<\infty,
\label{afftr}\end{equation}
where $q$ and $l$ are fixed parameters, it can be seen that $A=T^{-1}a$
is the raising operator for positive energy states.
In order to save space we shall describe
the whole hierarchy of such symmetry operators at once.
Let us introduce the operators
\begin{equation}
A=T^{-1}\prod_{k=1}^N\left({d\over dx}+\beta_k\right), \qquad
A^\dagger=\prod_{k=1}^N\left(-\;{d\over dx}+\beta_k\right)T,
\label{gen}\end{equation}
where $\beta_k$ are $N$ arbitrary real positive constants. 
It is easy to check that $A^\dagger$ and $A$ satisfy the following 
nonlinear algebraic relations
\begin{equation}
A^\dagger A=\prod_{k=1}^N\left(H+\beta_k^2\right),\qquad
A A^\dagger =\prod_{k=1}^N\left(q^2 H+\beta_k^2\right), \qquad
\label{algebra1}\end{equation}
\begin{equation}
HA^\dagger=q^2A^\dagger H, \qquad AH=q^2HA.
\label{algebra2}\end{equation}
For $N=1$ these relations define a $q$-analog of the Weyl algebra, or 
$q$-oscillator algebra (see, e.g. \cite{arik,macfarlane,kulish}) 
\begin{equation}
AA^\dagger-q^2A^\dagger A=\omega, \qquad \omega=\beta_1^2(1-q^2).
\label{qweylfree}\end{equation}
For $N=2$ one gets a $q$-analog of the $su(1,1)$ algebra in the form 
considered, e.g. in \cite{witten,flor}.
For $N>2$ one has polynomial quantum algebras 
\cite{s1,s2}. Note that in the limit $q\to 1$ one does not get
non-trivial algebras since the operators $A$ and $A^\dagger$ start to 
commute, still being the differential-difference operators for $l\neq 0$. 
For $q\neq 1$ it is possible to set $l=0$
by going to the reference frame where a fixed point of the affine
transformation is taken as the zero point, we shall assume this choice below.

Discrete series representations of the derived algebra are constructed by 
the action of the operators $A$ and $A^\dagger$ upon the Hamiltonian 
eigenstates  with $\lambda\neq 0$. The lowest weight series have the form
$$
A\psi_k^{(0)}(x)=0, \qquad \psi_k^{(0)}(x)\propto e^{-\beta_kx}, \qquad
k=1, 2, \dots, N,
$$
$$
\psi_k^{(n)}(x) \propto (A^\dagger)^n\psi_k^{(0)}(x), \qquad
H\psi_k^{(n)}(x)=-\beta_k^2q^{2n}\psi_k^{(n)}(x), \qquad n=0, 1, \dots, \infty.
$$
The wave functions $\psi_k^{(n)}(x)$ are not physical, their
eigenvalues accumulate near the $\lambda=0$ point from below. 
One can take the $\beta_k$ in (\ref{gen}) to be purely imaginary numbers. 
Then $\psi_k^{(n)}(x)$ describe continuous
spectrum states whose eigenvalues accumulate near the zero {\it from above}.
In this case $A^\dagger$ is not a hermitian conjugate of $A$ but 
still one can use representations of the algebra (\ref{algebra1}), 
(\ref{algebra2}). 
The highest weight representations appear as follows (in the same notations):
$$
A^\dagger\psi_k^{(0)}(x)=0, \qquad \psi_k^{(0)}(x)\propto e^{\beta_k q^{-1}x}, 
$$
$$
\psi_k^{(n)}(x) \propto A^n\psi_k^{(0)}(x), \qquad
H\psi_k^{(n)}(x)=-\beta_k^2q^{-2n-2}\psi_k^{(n)}(x).
$$
The eigenvalues of these states are unbounded from below
for real $\beta_k$ and they go to infinity for imaginary 
$\beta_k$. 
In any case operator $A$ is the lowering operator for negative
$\lambda$ eigenfunctions, 
but it raises the energy of continuous spectrum states. 

Physical  eigenstates of the Hamiltonian $H=-d^2/dx^2$ have the form 
$$
\psi_\lambda^\pm(x)={1\over\sqrt{2\pi}}\exp(\pm i\sqrt{\lambda} x), \qquad
$$
$$
\int_{-\infty}^\infty dx \; 
\psi^{\sigma *}_\lambda(x)\psi^{\sigma'}_{\lambda'}(x)=
\delta_{\sigma\sigma'}\delta(\sqrt{\lambda}-\sqrt{\lambda'}),
$$ 
where $\sigma, \sigma'=\pm$. 
The algebra generators act upon them in a simple way
\begin{equation}
A\psi_\lambda^\pm(x)=q^{-1/2}\prod_{k=1}^N\left(\pm i\sqrt\lambda+\beta_k\right)
\psi_{\lambda q^{-2}}^\pm(x),
\label{actiona}\end{equation}
\begin{equation}
A^\dagger\psi_\lambda^\pm(x)=q^{1/2}\prod_{k=1}^N
\left(\mp iq\sqrt\lambda+\beta_k\right)\psi_{\lambda q^{2}}^\pm(x).
\label{actionad}\end{equation}
We conclude that the free particle's Hilbert space provides
a unitary realization of the quite complicated symmetry algebras.
In the following sections we describe a generalization of this construction 
to nontrivial potentials along the lines of \cite{s1,s2}.

Consider coherent states of the above algebras defined as 
eigenstates of symmetry operators lowering the energy. Let us analyze first 
eigenstates of the ``annihilation" operator $A$, $A\psi_\alpha(x)=
\alpha\psi_\alpha(x)$. For the $q$-oscillator algebra 
such a definition has been considered, e.g.,  in 
\cite{arik,bied,chai,jurco,fivel2,alex,askey}
and many other recent papers. In our model these coherent 
states for $N=1$ are defined by the differential-delay equation
\begin{equation}
\psi_\alpha'(x)=\alpha\sqrt{q}\psi_\alpha(qx)-\beta_1\psi_\alpha(x),
\label{pant}\end{equation}
known in the literature as the   pantograph equation \cite{kato,iserles}.
Note that the initial value problem for this equation is highly nontrivial.
Using the results of the detailed analysis of (\ref{pant}) given in \cite{kato},
it is possible to see that solutions $\psi_\alpha(x)$ having finite
fixed values at $x=0$ are normalizable near the $x=\infty$ point when 
$|\alpha| < \beta_1$ but for $x\to-\infty$ they diverge 
exponentially fast. A similar situation holds for the general 
$N>1$ symmetry algebras.  We conclude that there are no physically acceptable 
coherent  states of this type in the free particle model. 
This is caused by the absence of a discrete spectrum. In the more 
complicated realizations of (\ref{algebra1}), (\ref{algebra2}) 
such states do exist. 

A different type of coherent states for the $q$-Weyl algebra 
has been constructed in \cite{s3}. 
These coherent states are defined as eigenstates of the operator $A^\dagger$,
which lowers the energy of the $\lambda > 0$ states, 
$A^\dagger\psi_\alpha(x)=\alpha\psi_\alpha(x).$
For the free particle realization of the $q$-oscillator algebra
these states are determined again by the pantograph equation but, now 
the dilation parameter is bigger than 1:
\begin{equation}
\psi_\alpha'(x)=-\alpha q^{-3/2}\psi_\alpha(q^{-1}x)+
\beta_1 q^{-1}\psi_\alpha(x).
\label{pantrue}\end{equation}
This time there are infinitely many normalizable functions $\psi_\alpha(x)$
for $|\alpha|>\beta_1$,  decreasing as 
$x^\kappa, \; q^\kappa=\alpha /q^{1/2}\beta_1$, for $|x| \to\infty$.
All of them have finite values at $x=0$ but they are not analytical near 
this point (i.e. Taylor series expansion does not converge).
For arbitrary $N$, coherent states of this type are defined by the generalized 
pantograph equation \cite{iserles}
\begin{equation}
\prod_{k=1}^N\left(-\; {d\over dx} + \beta_k\right)\psi_\alpha(qx)=
\alpha q^{-1/2}\psi_\alpha(x),
\label{pangen}\end{equation}
whose solutions are not expressible in terms of the classical special functions.
Expansion of normalizable solutions in the basis of 
Hamiltonian eigenfunctions is given by the integral 
\begin{equation}
\psi_{\alpha\pm}^{(s)}(x)={C(\alpha)\over\sqrt{4\pi}}\int_0^\infty 
{\lambda^{d_s}  e^{\pm i\sqrt{\lambda}x}d\lambda \over 
\prod_{k=1}^N(\pm iq\sqrt{\lambda}/\beta_k; q)_\infty}, 
\label{fourier}\end{equation}
$$
d_s={\ln \rho /\alpha q^{3/2} +2\pi is\over \ln q^2},
\qquad \rho\equiv\beta_1\cdots \beta_N,
$$
where $s=0, \pm 1, \dots$ is an integer enumerating linearly independent 
states and it is assumed that $0\leq \arg\alpha<2\pi$. In (\ref{fourier}) 
and below we use the standard notations for $q$-products \cite{gasper}
$$
(a; q)_\infty=\prod_{k=0}^\infty (1-aq^k), \qquad
(a; q)_n={(a; q)_\infty\over (aq^n; q)_\infty}.
$$
For positive integer $n$ one has
$$
(a; q)_n=\prod_{k=0}^{n-1}(1-aq^k), \qquad
(a; q)_{-n}={(-q/a)^n q^{n(n-1)/2}\over (q/a; q)_n}.
$$
The normalization constant $C(\alpha)$ is given by the integral 
\begin{equation}
|C(\alpha)|^{-2}=\int_0^\infty{ \lambda^{\tau} d\lambda\over 
\prod_{k=1}^N(-\lambda q^2/\beta_k^2; q^2)_\infty},\qquad
\tau = {\ln \rho /|\alpha|q \over \ln q}.
\label{normconst}\end{equation}
This constant is finite for $|\alpha|>\rho$, which is the
region of definition of the coherent states. 
For any $N$ the states (\ref{fourier}) 
have the common qualitative feature of non-analyticity near the $x=0$ point. 
The origin of this property  and possible physical consequences deserve 
further investigation.

One can look also for solutions of (\ref{pangen}) 
in the form of series similar to (\ref{cohfreeser}). 
For example, for $N=1$ one can write:
\begin{equation}
\psi_\alpha(x) \propto \sum_{n=-\infty}^\infty 
\left({\beta_1\sqrt{q}\over \alpha}\right)^n (i\theta q/\beta_1; q)_n 
e^{i\theta q^n x}, 
\label{contspec}\end{equation}
where $\theta$ is an arbitrary constant, $q<\theta\leq 1$. 
These non-normalizable solutions are 
bounded for any $x$ provided $|\alpha|>q^{1/2}\beta_1$, however it is not clear
whether they should be taken into account in order for coherent states to be 
complete.

It is worth mentioning that the pantograph equation appears in various 
problems. It has been encountered in a description of the light absorption by
interstellar matter \cite{ambart}, the 
collection of current by the pantograph of an electric
locomotive \cite{pant}, some number theory problem, etc (for a list of
applications, see \cite{iserles}). Here we have described another
physical application of this equation as the one determining the free particle
coherent states within the ladder-operator definition context. 

Two linearly independent $\lambda=0$ solutions of (\ref{freeeq})
can be represented in the form 
\begin{equation}
\psi_1(x)=1, \qquad \psi_2(x)=x+{q\over 1-q}\sum_{k=1}^N{1\over \beta_k}.
\label{zeroenergy}\end{equation}
Only the first one
is bounded and belongs to the continuous spectrum of the free particle. 
Formally upon $\psi_j(x)$, both 
operators $A$ and $A^\dagger$ are diagonalized simultaneously:
\begin{equation}
A\psi_j(x)=\rho q^{1/2-j}\psi_j(x), \qquad
A^\dagger\psi_j(x)=\rho q^{j-1/2}\psi_j(x).
\label{condens}\end{equation}
These are the simplest models of the $c$-number ``condensate"
representations of quantum algebras discussed in \cite{skorik,granov}. 

In a similar manner one can consider the free particle on the half-line. 
In order to make Hamiltonian self-adjoint it is necessary to impose boundary
conditions at $x=0$. The scaling operator defines physical symmetry only 
in the special cases $\psi(0)=0$ or   $\psi'(0)=0$ \cite{dhoker}. Then one
can define again coherent states as eigenstates of the lowering symmetry 
operators, but we shall not consider them here.

A curious fact is that for $N=1, \; q=i,\; \beta_1=1/\sqrt2$,  
one arrives at a simple differential-difference 
operator realization of the fermionic oscillator algebra. Indeed, it is 
not difficult to check that upon the states
\begin{equation}
A|0\rangle=0, \quad |0\rangle\propto e^{-x/\sqrt2}, \qquad
|1\rangle \equiv A^\dagger|0\rangle\propto e^{-ix/\sqrt2}
\label{fermion}\end{equation}
the relations $A^\dagger|1\rangle=0$ and $A|1\rangle=|0\rangle$ 
are satisfied, which means that 
$AA^\dagger+A^\dagger A=1, \; A^2=(A^\dagger)^2=0$. It is not clear whether
this formal two-dimensional representation has a unitary setting. 

The scaling operator $T$ also provides an interesting possibility 
of a non-standard realization of the ordinary bosonic oscillator algebra.
Let $Tf(x, y)=f(qx, qy)$ and $z=x+iy$, then the relations
\begin{equation}
[A, A^\dagger]=1, \qquad A=T^{-1}{d\over dz}, \qquad A^\dagger =zT,
\label{weylnew}\end{equation}
define some deformation of the Bargmann-Fock realization
(when $q$ is complex the operator $T$ scales and rotates $z$, we assume that
$0<q<1$). The measure density $\rho(|z|^2)$ in the scalar product
\begin{equation}
\langle\psi_1|\psi_2\rangle=\int d^2z \rho(|z|^2)\overline{\psi_1(z)} \psi_2(z)
\end{equation}
is found from the requirement for $A^\dagger $ to be a conjugate of $A$.
This gives the equation
\begin{equation}
{d\rho(t)\over dt}=- q^{-4}\rho\left({t\over q^2}\right), \qquad t=|z|^2,
\end{equation}
which we have just encountered in (\ref{cohfree}). Its solution
satisfying $\rho(0)=const$ can be represented in the form
$$
\rho(t)=\int_0^\infty dp\; h(p)e^{-pt}\exp{\ln^2 pq\over 4\ln q},\qquad
h(q^2p)=h(p).
$$
Acting by $A^\dagger$ upon the zero mode of $A$, $A\psi_0(z)=0,$
$\psi_0(z)=const$, one finds the Fock space basis vectors:
\begin{equation}
\psi_n(z)=C_n(q)z^n\propto (zT)^n 1.
\end{equation}
The normalization constants $C_n$ have the form
\begin{equation}
|C_n|^2={q^{n(n-1)}\over n! \sqrt{4\pi\ln 1/q} }
\left( \sum_{k=-\infty}^\infty (-1)^kh_k \exp{\pi^2k^2\over \ln q}
\right)^{-1},
\end{equation}
where $h_k$ are arbitrary constants appearing due to the nonuniqueness
of the measure. More precisely, $h_k$ are the coefficients of the Fourier 
expansion of an arbitrary periodic function entering the measure $h(p)$,
$$
h(p)=\sum_{k=-\infty}^\infty h_k \exp{\pi ik\ln p \over \ln q}.
$$
The relation between $h_k$ and $C_n$ shows 
that the moment problem for this measure
(i.e. determination of $\rho(t)$ from given $C_n$) does not have a
unique solution. Physical applications of the described realization of the 
Heisenberg-Weyl algebra are not known. 

\section{Coherent states of the $q$-deformed harmonic oscillator potential}
\setcounter{equation} 0

In section 2 we gave a simple derivation of the Yurke-Stoler
states (\ref{newcoh}) 
on the basis of a canonical transformation associated with the parity operator.
Actually, it was inspired by the 
analysis of coherent states for the $q$-oscillator algebra,  
\begin{equation}
AA^\dagger-q^2A^\dagger A=\omega, \qquad [A, \omega]=[A^\dagger, \omega]=0,
\label{qweyl}
\end{equation}
in the realization described in \cite{s1}. 
Let us consider this system in more detail. 
One can check that the pair of formal operators:
\begin{equation}
A=T^{-1}\left(d/dx+f(x)\right),\qquad 
A^\dagger=\left(-d/dx+f(x)\right)T,
\label{qlad}\end{equation}
where $T$ is the scaling operator (\ref{scaling}), 
satisfy (\ref{qweyl}) provided $f(x)$ is a solution of the equation
\begin{equation}
\frac{d}{dx}\left(f(x)+qf(qx)\right)+f^2(x)-q^2f^2(qx)=\omega,
\label{eq}
\end{equation}
derived in \cite{shabat} as the simplest self-similar
reduction of the dressing chain for Schr\"odinger equation.
The Hamiltonian of this system, 
$$ 
H= A^\dagger A-\nu= -d^2/dx^2+u(x), 
$$
\begin{equation}
u(x)=f^2(x)-f'(x)-\nu,\qquad \nu\equiv\omega/(1-q^2), 
\label{potential}
\end{equation}
satisfies the relations: 
\begin{equation}
AH=q^{2}HA, \qquad HA^\dagger=q^2A^\dagger H.
\label{qladder}
\end{equation}
For $q=1$ one has $f(x)=\omega x/2$, i.e. the standard harmonic oscillator. 
Suppose that $f(-x)=-f(x)$, which corresponds to the symmetric potential.
Then, in the limit $\omega \to 0$ the solution $f(x)$ analytical at $x=0$
vanishes due to the initial condition $f(0)=0$, 
and we get the zero potential model considered in the preceeding section 
with the factorization constant $E_0=0$ (the $E_0\neq 0$ case 
corresponds to the symmetric solution $f(x)=\sqrt{\nu}$). 

The analysis of \cite{skorik} shows that for complex values of $x$
and $q$, a solution of (\ref{eq}) analytical near $x=0$
exists and it is unique provided $|q|<1$ or $q$ is a
primitive root of unity of odd degree, $q^{2k+1}=1$. 
For $q^{2k}=1$ a solution may exist only for special 
initial conditions which, however, do not guarantee uniqueness. 
If $|q|=1$ but $q^n\neq 1$ then $f(x)=\pm\sqrt{\nu}$ are the only 
analytical solutions known to the author. For $0< |q|<1$ the function 
$f(x)$ cannot be
expressed in terms of known special functions. When $q^{2k+1}=1$, and in the
restricted case of $q^{2k}=1$, the problem is solved in terms of particular
hyperelliptic functions characterized by the presence of additional 
symmetries of the lattice of periods.
Since in the simplest cases, $q^3=1,\; q^4=1$, 
$q$ is just the modular parameter 
of elliptic functions, the function $f(x)$  (and its generalizations to be
described below) comprises hidden ``second" $q$-deformation
properties of hyperelliptic, or finite-gap potentials.

In \cite{sym} the system (\ref{qweyl})-(\ref{qladder}) has been derived from 
a special quantization of a simple model of classical mechanics (a particle 
in a finite depth potential) characterized by a quadratic Poisson algebra.
In this picture equation (\ref{eq}) has the form 
\begin{equation}
\hbar{d\over dx}\left( f(x)+ e^{\hbar\eta }f(e^{\hbar\eta }x)\right)
+f^2(x)-e^{2\hbar\eta }f^2(e^{\hbar\eta }x)=c(1-e^{2\hbar\eta }),
\label{qtrans}\end{equation} 
where $\hbar$ is Plank's constant, $q=e^{\hbar\eta }$, $\eta$ and 
$c$ are parameters of the classical potential.
Expanding this equation over $\hbar$ one finds successively
the classical, quasi-classical, and so on, approximations to the exact
solution $f(x)$ (note that in this approach the potential is an infinite
series over $\hbar$, there is a large nonuniqueness, etc). 

Let $x$ and $q^2$ be real, $\omega>0$, and $f(x)$ antisymmetric, $f(-x)=-f(x)$.
For $q^2>1$ the function $f(x)$ has singularities \cite{s1} so that
$A^\dagger$  is not conjugated to $A$ and the realization of 
(\ref{qweyl}) is not unitary. For $0<q^2<1$ the function $f(x)$
is bounded and has only one zero \cite{shabat}. These are the crucial
properties sufficient for $A^\dagger$ and $A$ to be well defined
operators in the Hilbert space. In particular, the zero mode of $A$, 
$A|0\rangle=0$, 
$\langle x|0\rangle \propto\exp (-\int^x f(y)dy)$, is normalizable and describes
the ground state of Hamiltonian $H$. As a result the whole spectrum of $H$ 
is found from unitary representations of the algebra (\ref{qweyl}).

It is not difficult to see from (\ref{qladder})
that the spectrum may consist of three parts: 
a discrete one, describing bound states
accumulating near the zero energy level from below, a 
continuous part going from zero to infinity and corresponding to 
scattering states, and, finally, a zero energy piece.
The discrete spectrum is described by the
lowest weight discrete series generated by $A^\dagger$ from the
vacuum state $|0\rangle$:
$$
|n\rangle=\frac{(A^\dagger)^n}{\sqrt{\omega^n [n]!}}|0\rangle, 
\qquad A|0\rangle=0,
\qquad \langle n | m\rangle=\delta_{nm},
$$
$$ 
[n]!=[n] [n-1]!,\quad [0]!=1, \quad [n]=(1-q^{2n})/(1-q^2),
$$
\vspace{1mm}
$$
A^\dagger |n\rangle=\omega^{1/2}\sqrt{1-q^{2(n+1)}
\over 1-q^2} |n+1\rangle, \qquad
A |n\rangle=\omega^{1/2}\sqrt{1-q^{2n}\over 1-q^2}|n-1\rangle, 
$$
\begin{equation}
H|n\rangle=E_n|n\rangle, \qquad E_n=-\; {\nu q^{2n}}.
\label{discrete}
\end{equation}
It  consists of one geometric series. Since zero modes of $A$ are
determined by the first order differential equation, it follows that
(\ref{discrete}) are 
the only physical states for $E<0$. Indeed, suppose that we missed one
physical state $|E\rangle$ with energy $E<0$. Acting by powers of $A$ upon
$|E\rangle$ we get a sequence of states of lower energies.
Since the potential is bounded this series should be truncated which is 
possible only if $|E\rangle$ is annihilated by some power of $A$, 
i.e. if the state $|E\rangle$ belongs to the series (\ref{discrete}).

The same argument shows that the states with positive energy, $E>0$, 
appear in the form of a geometric series infinite in both directions, 
\begin{eqnarray}
H|n\rangle_\lambda &=&\lambda q^{2n}|n\rangle_\lambda, 
\qquad n=0, \pm1, \pm2, \dots 
\label{cont} \\
A|n\rangle_\lambda &=& \sqrt{\nu+\lambda q^{2n}} |n-1\rangle_\lambda,
\nonumber \\
A^\dagger |n\rangle_\lambda &=& \sqrt{\nu+\lambda q^{2(n+1)}}
|n+1\rangle_\lambda,
\nonumber
\end{eqnarray}
where $\lambda>0$ is an arbitrarily chosen eigenvalue of $H$.
(This representation of the $q$-Weyl algebra was discussed in 
\cite{kulish,skorik}, cf. also \cite{rid}; 
note that it is not defined for $q\to 1$.)
Theoretically these states could be normalizable, but in our case this
is not so -- they belong to the continuous spectrum. Indeed, from 
(\ref{discrete}), (\ref{cont}) it follows that the scaled potential
$q^2u(qx)$ has the same spectrum as $u(x)$ except of the lowest state with
the energy $E_0=-\nu$. Similarly,  $q^{2k}u(q^k x)$ does not have the
$k$ lowest states. Taking the limit $k\to \infty$, one gets a system without 
a negative spectrum, whereas the states (\ref{cont}) are not washed away.
From the boundedness of the initial potential it follows that 
$q^{2k}u(q^k x)\to 0$ for $k\to \infty$ because $q^2<1$. This means that
our potential is reflectionless, being obtained by a special infinite
step dressing of zero potential. But for zero potential 
the series (\ref{cont}) with $\omega=0$ (removal of the levels is 
equivalent to rescaling $\omega\to q^{2k}\omega\to 0$) correspond to 
the continuous spectrum (see the preceeding section). A rigorous 
proof of the absence of positive energy bound states 
requires an estimate of the asymptotics of the potential \cite{reed}.
The $|x|\to\infty$ asymptotics of (\ref{potential})
proposed in \cite{degas} decreases sufficiently fast  
in order to guarantee the absence of such exotic states, 
$u(x) \to h(x)/x^2+O(1/x^3)$, where $h(qx)=h(x)$ is a bounded function.

Upon the zero modes of Hamiltonian the operators $A$ and $A^\dagger$
behave like integrals of motion, i.e. they commute with $H$. Since the 
relations $A^\dagger A=A A^\dagger=\nu$ can be satisfied by arbitrary
invertable matrix, the dimension of this representation is not restricted.
It may be either infinite-dimensional under additional requirements
\cite{kulish} or just one-dimensional. In the latter case 
the creation and annihilation operators degenerate
into complex numbers \cite{skorik,granov},
\begin{equation}
H|cl\rangle=0, \qquad A|cl\rangle=\sqrt{\nu} e^{-i\theta}|cl\rangle,
\qquad A^\dagger|cl\rangle=\sqrt{\nu} e^{i\theta}|cl\rangle,
\label{zeromodes}
\end{equation}
where we assume that the state $|cl\rangle$ is normalizable.
Note that the limit $q\to 1$ is not defined.
In our case these ``classical" states correspond to the boundary
between discrete and continuous spectrum. 
As we shall show below, in the $q$-oscillator model (\ref{qlad})
with $f(0)=0$ the corresponding wave functions are not bounded
and their eigenvalues differ from those in (\ref{zeromodes}).

{\it Remark.} Actually, the existence of the non-zero $c$-number 
representations is not very rare for quantum algebras. For example, for 
the Cartesian version of the $sl_q(2)$ algebra 
$$
q^{-1}J_1J_2 - qJ_2J_1= J_3, \quad
q^{-1}J_2J_3 - qJ_3J_2= J_1, \quad
q^{-1}J_3J_1 - qJ_1J_3= J_2,
$$ 
describing dynamical symmetries of some discrete reflectionless potentials
\cite{spirizhed},
one can set $J_k=1/(q^{-1}-q)$ and the algebra is satisfied. 
Such exotic representations are often skipped in discussions 
of applications of quantum algebras. 
\smallskip

As we have discussed already, coherent states of the first type of the algebra 
(\ref{qweyl}) are defined as eigenfunctions of the annihilation operator $A$. 
These states are built from the lowest weight discrete series 
representation (\ref{discrete}) (see, e.g. \cite{arik}):
\begin{equation}
A|\alpha, q\rangle =\alpha|\alpha, q\rangle, \qquad 
|\alpha, q\rangle=C(\alpha)
\sum_{n=0}^\infty\frac{\alpha^n}{\sqrt{\omega^n [n]!}}|n\rangle, 
\label{qcoh}
\end{equation}
where
$$
|C(\alpha)|^{-2}\equiv e_{q^2}(z)=\sum_{n=0}^\infty\frac{z^n}{(q^2; q^2)_n}=
{1\over (z; q^2)_\infty}, \qquad z= {|\alpha|^2(1-q^2)\over\omega},
$$
is a $q$-analog of the exponential function, or $_1\varphi_0(0; q^2, z)$
basic hypergeometric function. The general $q$-series of such type 
is defined as follows \cite{gasper}:
$$
_r\varphi_s \Biggl( {a_1, \; a_2,\; \dots, \; a_r \atop
b_1,\; b_2,\; \dots, \; b_s}; q, z\Biggr)=
\sum_{n=0}^\infty {(a_1; q)_n (a_2; q)_n \dots (a_r; q)_n\over
(q; q)_n (b_1; q)_n \dots (b_s; q)_n}[(-1)^nq^{n(n-1)/2}]^{1+s-r} z^n,
$$
where $r$ and $s$ are arbitrary positive integers, and $a_1, \dots, a_r,
b_1, \dots, b_s$ are free parameters.
The states (\ref{qcoh}) are normalizable only for $|\alpha|^2< \nu$.
Since the Hilbert space of our model is larger than 
the discrete spectrum Fock space spanned by $|n\rangle$,
the states (\ref{qcoh}) {\it are not complete}. There remain 
two parts corresponding to zero and positive energy eigenvalues.  
Although the latter are not normalizable, one cannot discard them. 
Expansion of the eigenfunctions of $A$ over the fixed energy states
should contain in general an integral over the continuous spectrum. 
This situation differs drastically from the $q=1$ algebra case 
where the Fock space was complete.  

Note that the definition (\ref{qcoh}) works for $q^2>1$ as well since 
$A$ remains to be the lowering operator for positive energy states. Then, the 
states $|\alpha\rangle$ are normalizable for arbitrary value of $\alpha$,
\begin{equation}
|C(\alpha)|^{-2}\equiv E_{q^{-2}}(z)=\sum_{n=0}^\infty {z^nq^{-n(n-1)}\over
(q^{-2}; q^{-2})_n}=(-z; q^{-2})_\infty, \qquad 
z={|\alpha|^2(1-q^{-2})\over\omega},
\label{newnorm}\end{equation}
where $E_{q^{-2}}(z)$ is another analog of the exponential function,
or $_0\varphi_0(q^{-2}, -z)$ basic hypergeometric function.
In this case coherent states are complete because the
Hamiltonian has only a discrete spectrum as in \cite{arik}. 

For $q^2<1$, coherent states formed by the positive energy states
should be defined as eigenstates of the operator $A^\dagger$, 
\begin{equation}
A^\dagger|\alpha, q\rangle =\alpha|\alpha, q\rangle,
\label{qcohplus}\end{equation}
since for $E>0$, not $A$, but $A^\dagger$ lowers the energy \cite{skorik}.
Suppose that for some $\lambda$ the states of the series (\ref{cont})
are normalizable (this is not so in  our case, but in principal it is
possible). Then we find
\begin{equation}
|\alpha, q\rangle_\lambda = C(\alpha) \left(
|0\rangle_\lambda + \sum_{n=1}^\infty
\left( 
{\nu^{n/2}\over  \alpha^n} (-\lambda q^2/\nu; q^2)_n^{1/2}
|n\rangle_\lambda +
{\alpha^n q^{n(n-1)/2}\over \lambda^{n/2}(-\nu/\lambda; q^2)_n^{1/2}}
|-n\rangle_\lambda
\right)\right). 
\label{newqcoh}
\end{equation}
The normalization constant $C(\alpha)$ is related to the bilateral
basic hypergeometric series $_0\psi_1$,
$$
|C(\alpha)|^{-2} =  {_0\psi_1}(b ; q^2, z), 
\qquad b=-\nu/\lambda, \quad z=-|\alpha|^2/\lambda.
$$
The general bilateral $q$-hypergeometric series is defined as follows 
\cite{gasper}:
$$
_r\psi_s\left({a_1, \dots, a_r \atop b_1, \dots, b_s}; q, z\right)=
\sum_{n=-\infty}^\infty {(a_1; q)_n \dots (a_r; q)_n \over
(b_1; q)_n \dots (b_s; q)_n} \left((-1)^nq^{n(n-1)/2}\right)^{s-r}z^n.  
$$
Using the Ramanujan sum for the $_1\psi_1$ series \cite{gasper}
one can express $C(\alpha)$ in terms of the infinite products
$$
|C(\alpha)|^{-2} = 
{(q^2; q^2)_\infty(-\lambda q^2/|\alpha|^2; q^2)_\infty 
(-|\alpha|^2/\lambda; q^2)_\infty
\over (-\nu/\lambda; q^2)_\infty (\nu/|\alpha|^2; q^2)_\infty}.
$$
The states (\ref{newqcoh}) are normalizable when 
$|\alpha|^2>\nu$, i.e. $\alpha$ should lie outside
of the region where the states (\ref{qcoh}) were defined.

When positive eigenvalues of the abstract Hamiltonian $H$,
$H|\lambda\rangle=\lambda|\lambda\rangle$, form a 
continuous spectrum, the states $|\alpha, q\rangle$ are defined by 
the integral over $|\lambda\rangle$, 
\begin{equation}
|\alpha, q\rangle = \int_0^\infty d\lambda\;
{\lambda^d \chi(\lambda, \alpha)\; |\lambda\rangle \over 
\sqrt{(-\lambda q^2/\nu; q^2)_\infty}}, 
\qquad d={\ln \sqrt{\nu}/q^2\alpha\over \ln q^2},
\label{newqcohcont}
\end{equation}
where $\chi(\lambda)$ is an arbitrary function periodic on the logarithmic
scale, $\chi(q^2\lambda)=\chi(\lambda)$. In principle the continuous 
spectrum may have an infinity of gaps, each type of gap appearing in the form
of geometric series. Here we assume that it fills the whole
interval $0<\lambda<\infty$. From the requirement for $A$ and 
$A^\dagger$ to be hermitian conjugates of each other, with the action
(which is defined only up to an arbitrary phase factor depending on $\lambda$)
$$
A|\lambda\rangle =\sqrt{\nu+\lambda}\;|\lambda q^{-2}\rangle,\qquad
A^\dagger |\lambda\rangle = \sqrt{\nu+\lambda q^2}\;|\lambda q^2\rangle
$$ 
one finds the form of the scalar product
\begin{equation}
\langle \lambda|\sigma\rangle=\lambda\delta(\lambda-\sigma).
\label{scalarproduct}\end{equation}
In our case there are two sets of states (\ref{newqcohcont})
because the continuous spectrum is doubly degenerate. 
Expanding the function $\chi(\lambda)$ into a Fourier series, we obtain 
an infinite number of linearly independent coherent states of the form
\begin{equation}
|\alpha, q\rangle_{s}=C(\alpha)\int_0^\infty 
{d\lambda \; \lambda^{d_s}\; |\lambda\rangle\over 
(-\lambda q^2/\nu; q^2)^{1/2}_\infty},\qquad 
d_s=d+{2\pi i s\over \ln q^2},\quad s=0, \pm 1, \dots, 
\label{cohfourier}
\end{equation}
which are normalizable for $|\alpha|^2>\nu$. The normalization 
constant is calculated exactly being determined by the special case of the 
Ramanujan $q$-beta integral \cite{gasper}
$$
|C(\alpha)|^{-2}=\int_0^\infty
{d\lambda\; \lambda^\tau \over (-\lambda q^2/\nu; q^2)_\infty}
=-\;{\pi \over \sin{\pi\tau}}
{(q^{-2\tau}; q^2)_\infty \over (q^2; q^2)_\infty}
\left({\nu\over q^2}\right)^{\tau+1},
$$
where $\tau=2 Re\; d+1$. The fact that there exists an infinity of normalizable 
states (\ref{cohfourier}) leads to an interesting effect in the model of
$q$-oscillator interacting with a classical current \cite{lsz}.

Consider the structure of coherent states of the first type (\ref{qcoh}) in 
the realization (\ref{qlad}), (\ref{eq}).
Denote $\psi_\alpha(x)=\langle x|\alpha, q\rangle$. These wave
functions are defined by the equation:
\begin{equation}
(d/dx+f(x))\psi_\alpha(x)=\alpha\sqrt{|q|}\psi_\alpha(qx),
\label{defqcoh}
\end{equation}
where $f(x)$ is a solution of (\ref{eq}). Unfortunately, a full
description of the properties of $\psi_\alpha(x)$ is not accessible at 
present. Nevertheless, several crucial points can be brought into view. 

Suppose $0<q<1$; then in the $x\to\infty$ limit we get the equation
\begin{equation}
\psi_\alpha'(x)=\alpha\sqrt{q}\psi_\alpha(qx)-\sqrt{\nu}\psi_\alpha(x),
\label{eqinfty}\end{equation}
which is exactly the pantograph equation encountered in the free
particle model. 
Using the corresponding analysis, we conclude that the $x\to\infty$
asymptotics of coherent states is
\begin{equation}
\psi_\alpha(x)={h(x, \alpha)\over x^\kappa}, \qquad 
\kappa={\ln\alpha\sqrt{q/\nu}\over \ln q},
\label{asympt}\end{equation}
where $h(x)$ is some function satisfying $h(qx)=h(x)$.
A similar asymptotics of $\psi_\alpha(x)$ holds for $x\to-\infty$
because in this limit $f(x)\to -\sqrt{\nu}$. Therefore coherent states are 
normalizable near infinity provided $Re\; \kappa>1/2$, or $|\alpha|^2<\nu$, 
consistent with the previous considerations. Due to the arbitrariness of 
$h(x, \alpha)$, asymptotics (\ref{asympt}) corresponds to a countable set of 
solutions of (\ref{eqinfty}) for fixed $\alpha$,  
whereas it is possible to construct 
only one coherent state by superposing discrete spectrum states.
Probably it is the requirement for analyticity at $x=0$ that
determines the latter solution uniquely. Then the non-analytical
solutions should not be normalizable  near this point. 
For $Re\; \kappa<0$ the functions 
$\psi_\alpha(x)$ are not bounded, and so not physical. 
The significance of the bounded at infinity but
not normalizable solutions of (\ref{defqcoh}), appearing in the range 
$0\leq Re\; \kappa \leq 1/2$, is not yet clear. 
This problem is related to the analysis of the completeness of
coherent states which is beyond the scope of the present work.

Consider the zero modes of the Hamiltonian for the solution of (\ref{eq}) 
determined by  the 
condition $f(0)=0$. Denoting $\psi_{cl}(x)=\langle x|cl\rangle$, we have
\begin{equation}
\psi_{cl}''(x)=u(x)\psi_{cl}(x).
\label{clfunct}\end{equation}
The first several terms of the Taylor expansion of the functions $f(x)$ 
and $u(x)$ are easily found:
$$
f(x)={\omega  x\over 1+q^2}+ {(q^2-1)\omega^2 x^3\over 3(1+q^2)(1+q^4)}
+ O(x^5),
$$
$$
u(x)={2\omega \over q^4-1}+{2\omega^2 x^2\over (1+q^2)^2(1+q^4)}+O(x^4).
$$
Looking for $\psi_{cl}(x)$ in the form of a Taylor series, we find the first
three terms of the  odd and even wave functions:
$$
\psi_{cl}^{even}(x)=1 + {\omega x^2\over q^4-1} + {(1-q^2+q^4)\omega^2x^4\over 
3(1+q^4)(1-q^4)^2} + O(x^6),
$$
$$
\psi_{cl}^{odd}(x)=x+{\omega x^3 \over 3(q^4-1)} + 
{(2-3q^2+2q^4)\omega^2x^5 \over 15(1+q^4)(1-q^4)^2} + O(x^7).
$$
Since the operator $A$ maps the space of zero modes onto itself, there
should be at least one eigenfunction of $A$. Taking the linear combination
of the above solutions we find that there are two such functions 
\begin{equation}
A\psi_{cl}^{\pm}=\pm i\sqrt{\nu}q^{-1}\psi_{cl}^{\pm},\qquad
\psi_{cl}^{\pm}(x)=\psi_{cl}^{even}(x)
\pm i\sqrt{\nu}q^{-1/2}\psi_{cl}^{odd}(x).
\label{cleigen}\end{equation}
Evidently $\psi_{cl}^{\pm}$ are eigenfunctions of the operator $A^\dagger$ 
as well, $A^\dagger\psi_{cl}^{\pm}=\mp iq\sqrt{\nu}\psi_{cl}^{\pm}$. 
The eigenvalues of $A$ and $A^\dagger$ are not complex conjugate to each other.
Hence, the ``condensate" representations of the $q$-oscillator
algebra do not belong to the discrete spectrum, actually, $\psi_{cl}^{\pm}$ are
not even bounded. 
This can be verified by the direct estimation of the asymptotics of these
functions using the fact that they satisfy equation (\ref{defqcoh})
with $\alpha=\pm i \sqrt{\nu}/q$. We, however, choose a slightly 
different approach. Consider the eigenvalue equation for $A^\dagger$
\begin{equation}
(-d/dx+f(x))\sqrt{q}\psi_{cl}^\pm(qx)=\mp iq\sqrt{\nu}\psi_{cl}^\pm(x).
\label{eqcl}
\end{equation}
Eliminating the derivative part from (\ref{defqcoh}) and (\ref{eqcl}), we get:
\begin{equation}
(f(x)+q^{-1}f(q^{-1}x))\psi_{cl}^\pm(x)=
\pm i\sqrt{\nu}q^{-1/2}\left(\psi_{cl}^{\pm}(qx)-\psi_{cl}^\pm(q^{-1}x)\right),
\label{dse}
\end{equation}
i.e. zero energy wave functions satisfy the second order purely 
finite-difference equation. 
Since for $x\to \infty$, $f(x)\to \sqrt{\nu}$, the leading asymptotics 
of $\psi_{cl}^\pm$ are found as solutions of the {\it free} finite-difference 
Schr\"odinger equation, which are
\begin{equation}
\psi_{cl}^\pm(x) = \chi^\pm(x)e^{ik_\pm\ln x},\qquad \chi^\pm(qx)=\chi^\pm(x). 
\label{clasymp}\end{equation}
Substituting this Ansatz into (\ref{eqinfty}) with $\alpha=\pm i\sqrt{\nu}/q$,
one finds $2ik_\pm=1 \mp{i\pi/\ln q}.$
The derived asymptotics of $\psi_{cl}^\pm(x)$ 
look like Bloch wave functions for a particle in a periodic  
potential with coordinate variable being $\ln x$.
Note that the quasimomenta $k_\pm$ are complex, which forces the 
wave function to increase at infinity as $\sqrt{x}$, unlike the $u(x)=0$ case
when a symmetric constant wave function was bounded. A more detailed 
consideration  of such non-standard implementation of the Bloch's theorem is 
given in section 8. 

Let us discuss briefly the $-1<q<0$ case. Due to the parity symmetry
one has the relation: $A^2\big|_{q<0}=-A^2\big|_{q>0}$. It means that the 
$q>0$ coherent states $|i\alpha, q\rangle$ and $|-i\alpha, q\rangle$
provide independent eigenstates of $A^2\big|_{q<0}$. 
The parity transformation changes only the sign of $\alpha$. 
Therefore, the eigenstates of $A$ for $q<0$
are given by the parity coherent states (\ref{newcoh}) where
on the  r.h.s. one has $q$-coherent states for $q>0$. Moreover, the 
precise realization (\ref{newladder}) and  superposition (\ref{newcoh}) 
appear in the $q\to -1$ limit of the $q$-oscillator system 
under consideration \cite{skorik}. 
Indeed, the general solution of equation (\ref{eq}) for $q=-1$ is 
$f(x)=\omega x/2$, and the variation of $q$ from $1$ to $-1$ performs a 
transition from the canonical coherent states to the parity
coherent states (\ref{newcoh}). Note however, that if one takes solution
of (\ref{eq}) satisfying asymmetric initial condition $f(0)\neq 0$,  
then the $q\to -1$ limit simply does not exist and for $-1< q<0$ the parity
operator does not help in the analysis.

Coherent states of the second type are defined as eigenstates 
of the $A^\dagger$ operator (\ref{qcohplus}):
\begin{equation}
(-d/dx+f(x))\sqrt{q}\psi_\alpha(qx)=\alpha\psi_\alpha(x),
\label{cohqweyl}\end{equation}
where we use the same notations as in (\ref{defqcoh}). 
In the $x\to\infty$ limit one gets again the pantograph
equation, but with the scaling parameter  $q^{-1}$:
\begin{equation}
\psi_\alpha'(x)=-\alpha q^{-3/2}\psi_\alpha(q^{-1}x)+ q^{-1}\sqrt{\nu}\psi(x).
\label{cohqweylas}\end{equation}
From the preceding section we know that this equation has solutions
with asymptotics
$$
\psi_\alpha(x)\to {h(x, \alpha)\over x^\kappa}, \qquad 
h(qx)=h(x), \quad \kappa={\ln\sqrt{\nu q}/\alpha \over \ln q}.
$$
These coherent states are normalizable provided $Re\; \kappa>1/2$, or
$|\alpha|^2>\nu$. Their expansion over the continuous spectrum states
was presented in the abstract form (\ref{cohfourier}). 
For $Re\;\kappa < 0$, or $|\alpha|^2<q\nu$ the 
functions $\psi_\alpha(x)$ are not bounded. For
$0\leq Re\;\kappa \leq 1/2$ we get again wave functions that are bounded at 
$|x|\to \infty$ but unnormalizable. It looks like they do not have an 
expansion over the 
eigenstates of the Hamiltonian and the reason for this needs clarification. 
In the considered models there are no normalizable coherent 
states for the circle $|\alpha|^2=\nu$. In principle it is possible that
zero modes of a Hamiltonian $H$ are normalizable, in which case it is
natural to count them among the coherent states (otherwise there will be 
no completeness). 

\section{General class of self-similar 
potentials and their coherent states}
\setcounter{equation} 0

It is well known \cite{abl} that the one-dimensional Schr\"odinger equation
\begin{equation}
H\psi(x)=-\psi^{\prime\prime}(x) + u(x)\psi(x) = \lambda \psi(x),
\label{a}
\end{equation}
has an important non-quantum mechanical application 
in the theory of non-linear evolution equations. In particular, the 
Korteweg-de Vries (KdV) equation can be solved with the help
of the inverse scattering method for two classes of initial conditions
$u(x,t=0)$. The first one consists of the potentials $u(x)$ satisfying
the restriction $\int_{-\infty}^\infty (1+|x|)|u(x)|dx<\infty$, which
guarantees that the number of bound states is finite.
Reflectionless potentials with $N$ discrete eigenvalues
are the simplest examples from this family.
Since they generate $N$-soliton solutions of the 
KdV equation they are called the soliton potentials. 
The second class is related to non-singular periodic (or quasiperiodic) 
functions, 
$u(x+ l)=u(x),$ characterized by the presence of $N$ gaps of finite width
in the spectrum of (\ref{a}). These finite-gap (hyperelliptic) potentials
\cite{dubrovin} are reduced to the solitonic ones in the limit $l \to\infty$. 
They can be thought of as superpositions of an infinite number of solitons 
(``periodic solitons"), but there is no scattering problem for such 
objects, i.e. the solitary character of the ingredient waves is lost.
There is a third relatively simple class of self-similar solutions of the KdV 
equation 
related to the Painlev\'e transcendents requiring a different treatment.

Recently, reflectionless potentials with an infinite number of 
discrete levels have been systematically considered in 
\cite{march,shabat,s1,duke,novok,skorik,degas}. 
For $x\to \infty$ such potentials decrease slowly and the standard
inverse scattering method does not produce constructive results.
These potentials deserve to be named the infinite soliton ones, since
they do not reflect and may be approximated with some accuracy by 
the $N$-soliton potentials $(N < \infty)$. Moreover this class absorbs
the finite-gap potentials which emerge for 
special limiting values of parameters. The related problem of 
the approximation of confining and band spectrum potentials with
the help of reflectionless potentials was discussed earlier in 
\cite{ros1,ros2}. 

Let us describe briefly the factorization method \cite{infeld,miller} 
that allows us to find the  particular subclass of infinite-soliton potentials 
characterized by the $q$-deformed symmetry algebras. 
This method was invented in quantum mechanics by Schr\"odinger, 
it is deeply related to the Darboux (B\"acklund, dressing, etc) 
transformations for linear differential equations.  
Within this approach, one takes a set of Hamiltonians,
\begin{equation}
L_j=-d^2/dx^2+u_j(x),\qquad j=0, \pm1, \pm2, \dots
\label{hamj}\end{equation}
and represents them as products of the 
first-order differential operators, 
\begin{equation}
A_j^+=-\; {d\over dx}+f_j(x),\qquad A_j^-={d\over dx}+f_j(x),
\label{e2}
\end{equation}
up to some constants $\lambda_j$:
\begin{equation}
L_j=A_j^+A_j^- + \lambda_j,
\label{e3}
\end{equation}
i.e. $u_j(x)=f_j^2(x)-f'_j(x)+\lambda_j$. Then one imposes the following
intertwining relations:
\begin{equation}
L_jA^+_j=A_j^+L_{j+1},\qquad A_j^-L_j=L_{j+1}A_j^-, 
\label{e4}
\end{equation}
which constrain the difference in spectral properties of $L_j$ and 
$L_{j+1}$ and are equivalent to the equations:
\begin{equation}
A^+_{j+1}A^-_{j+1}+\lambda_{j+1}=A^-_jA^+_j+\lambda_j.
\label{e5}
\end{equation}
The same results are obtained if one starts from the linear equations
\begin{equation}
L_j\psi_j=\lambda\psi_j, \qquad \psi_{j+1}=A^-_j\psi_j.
\label{e5a}
\end{equation}
The compatibility condition of (\ref{e5a}) is given by (\ref{e4}). 
Resolving the latter relations, one finds $\lambda_j$ as the integration 
constants such that the factorization (\ref{e3}) takes place.
These are the basic ingredients of the factorization method which allows us
to construct new solvable Schr\"odinger equations from  a given one.

Substitution of (\ref{e2}) into  (\ref{e5}) yields the chain of
differential equations:
\begin{equation}
f'_j(x)+f'_{j+1}(x)+f^2_j(x)-f_{j+1}^2(x)=\mu_j, \quad 
\mu_j \equiv \lambda_{j+1}-\lambda_j,
\label{e6}
\end{equation}
which is called the dressing chain \cite{shabat,vs}. 
Any spectral problem with a known nontrivial discrete spectrum 
generates some solution of (\ref{e6})
such that the ordered discrete eigenvalues are given by the 
constants $\lambda_j$. 
The factorization method works with the inverse problem  -- 
to find such solutions of (\ref{e5}) or (\ref{e6})
for which the spectrum of the associated Schr\"odinger operator
will be determined automatically. Note that in some cases the spectrum
can be found even if $\lambda_j$ do not belong to it,
e.g., such a situation can take place for hyperelliptic potentials 
\cite{dubrovin}, \cite{weiss}-\cite{skorik2}. 

The variable $j$ was playing above the role of a label. It could therefore 
be removed in favor of other notations, e.g.
$f_j\equiv f, \; f_{j+1}\equiv\tilde f$, etc. However, it is convenient 
to think of $j$ as a discrete set of points on a continuous manifold.
Then one can consider $j$ (or its function) 
as a continuous variable and look for solutions of
(\ref{e6}) in series form $f_j(x)=\sum g_k(x) j^k$. 
Infeld and Hull \cite{infeld} have considered
such an Ansatz and have found that the series contains a finite number of
terms iff $f_j(x)=\alpha(x)j+\beta(x)+\gamma(x)/j$, where 
$\alpha, \beta, \gamma$ are some elementary functions. 
This does not mean that the infinite-series solutions are meaningless,
it indicates rather that truncation of the series is related to 
some simple symmetries \cite{miller}. 
It is difficult to work with formal power series, 
often even their convergence is not known. Analysis of
solutions characterized by separation of variables or by some special 
dependence on them is essentially simplified. 
In the Lie theory of differential equations solutions of such type
are called the similarity, or self-similarity solutions.
Unfortunately for the differential-difference equations there is no
complete theory of such solutions; a subclass of them 
can be found using the methods developed for purely differential
equations \cite{levi}. An additional 
complication associated with equation (\ref{e6}) consists of the 
fact that there are two unknowns, $f_j(x)$ and $\mu_j$, i.e. the 
system is highly underdetermined.

According to definition, (self-)similarity solutions are the 
solutions invariant under symmetry transformations of a given equation.
The potentials we are interested in appear as fixed points of the 
combination of an affine transformation of the coordinate,
$x\to qx+l,$ and a shift along the discrete lattice, $j\to j+N$.
Indeed, the change in numeration of solutions by an integer, maps
solutions of (\ref{e6}) to the solutions, $f_j(x)\to f_{j+N}(x),\; 
\mu_j\to \mu_{j+N}$. The same is true for the affine group,
$f_j(x)\to qf_j(qx+ l),\; \mu_j\to q^2\mu_j$. We may look for the 
class of solutions invariant under both these symmetry transformations:
\begin{equation}
f_{j+N}(x)=qf_j(qx+ l),\qquad \mu_{j+N}=q^2\mu_j.
\label{e8}
\end{equation}
These relations define a class of potentials that we are going 
to analyze below. The simplest reduction of such type, 
$f_j(x)=q^jf(q^jx),$ $\lambda_j=q^{2j}$, 
corresponding to $N=1,\; l=0$ in (\ref{e8}), has been 
found by Shabat \cite{shabat}.
Actually, the general class of closures of the dressing 
chain (\ref{e8}) has been introduced in a way different from the above 
\cite{s2}, namely, from $q$-deformation of the 
parasupersymmetric quantum mechanics based upon some 
polynomial algebras \cite{rubakov}. 

At the operator level, the relations (\ref{e8}) lead to the Schr\"odinger
operators with non-trivial $q$-deformed symmetry algebras.
Let us consider the products:
\begin{equation}
M_j^+=A_j^+A^+_{j+1}\dots A^+_{j+N-1},\qquad
M_j^-=A_{j+N-1}^-\dots A^-_{j+1}A_j^-,
\label{e9}
\end{equation}
which generate the intertwinings
\begin{equation}
L_jM_j^+=M_j^+L_{j+N},\qquad M_j^-L_j=L_{j+N}M_j^-.
\label{e10}
\end{equation}
The structure relations complimentary to (\ref{e10}) appear as 
\begin{equation}
M_j^+M_j^-=\prod_{k=0}^{N-1}(L_j-\lambda_{j+k}),\qquad
M_j^-M_j^+=\prod_{k=0}^{N-1}(L_{j+N}-\lambda_{j+k}).
\label{e11}
\end{equation}
Equations (\ref{e10}), (\ref{e11}) can be rewritten as the higher order
polynomial supersymmetry algebra \cite{and}:
\begin{equation}
\{Q^+, Q^-\}=\prod_{k=0}^{N-1}(K-\lambda_k), \qquad
\left( Q^\pm\right)^2=[K, Q^\pm]=0,
\label{susy}
\end{equation}
where 
$$
Q^+=\left(\matrix{0&M^+_0\cr 0&0\cr}\right), \qquad
Q^-=\left(\matrix{0&0\cr M^-_0 &0\cr}\right), \qquad
K=\left(\matrix{L_0&0\cr 0&L_N\cr}\right),
$$
the index $j=0$ was fixed for simplicity of notations. 
Presence of nonlinearity brings essentially new features with respect 
to the standard supersymmetric quantum mechanics case corresponding to $N=1$.

The identities (\ref{e10}) show that if the operators 
$L_j$ and $L_{j+N}$ are
related to each other through some simple similarity transformation, e.g.
\begin{equation}
L_{j+N}=q^2 TL_jT^{-1}+\omega,
\label{e12}
\end{equation}
where $T$ is some invertable operator, then the combinations
$$
B_j^+\equiv M_j^+ T,\qquad B_j^-\equiv T^{-1}M_j^-,
$$ 
map eigenfunctions
of $L_j$ onto themselves, i.e. they describe symmetries of $L_j$.
The form of $T$ is restricted by the requirement for the $L_j$'s to be
of the Schr\"odinger form. The closure (\ref{e8}) corresponds to the 
choice of $T$ as the affine transformation operator, 
$Tf(x)=\sqrt{|q|}f(qx+l)$. Fixing the indices and 
removing their irrelevant part ($L\equiv L_0, \; B^\pm\equiv B^\pm_0$), 
we get the symmetry algebra \cite{s1,s2}:
\begin{equation}
LB^+ - q^2B^+L=\omega B^+,\qquad B^-L - q^2 L B^- =\omega B^-,
\label{e13}
\end{equation}
\begin{equation}
B^+B^-=\prod_{k=0}^{N-1}(L-\lambda_k),\qquad
B^-B^+=\prod_{k=0}^{N-1}(q^2L+\omega-\lambda_k).
\label{e14}
\end{equation}
After the shift of the zero energy point, 
$$
H\equiv L-{\omega \over 1-q^2},\qquad E_k\equiv \lambda_k -
{\omega \over 1-q^2}, 
$$
the algebra takes a simpler form 
\begin{eqnarray}
HB^\pm &=&q^{\pm2}B^\pm H, \quad \nonumber \\
B^+B^- &=&\prod_{k=0}^{N-1}(H-E_k),\quad
\label{goodalgebra} \\
B^-B^+ &=& \prod_{k=0}^{N-1}(q^2H-E_k). 
\nonumber
\end{eqnarray}
This algebra was already met in the discussion of coherent states
of the free nonrelativistic particle. 

Let us write out explicitly the system of nonlinear differential 
equations with deviating argument that one needs to solve in order 
to find the explicit form of the self-similar potentials:
\begin{eqnarray}
 {d\over dx}\left(f_0(x)+f_1(x)\right)+f_0^2(x)- f_1^2(x)&=&\mu_0,
\nonumber \\
{d\over dx}\left(f_1(x)+f_2(x)\right)+f_1^2(x)- f_2^2(x)&=&\mu_1, 
\nonumber \\  
\dots \dots \dots \nonumber \\
{d\over dx}\left(f_{N-1}(x)+qf_0(qx+ l)\right)+f_{N-1}^2(x)- 
q^2f_0^2(qx+l)&=&\mu_{N-1}.  
 \label{e15} 
\end{eqnarray}
Note that the limit $q\to 1$ is not trivial. For non-zero parameter $l$ 
we get a realization of the algebra  (\ref{e13}), 
(\ref{e14}) at $q=1$ which generalizes the one described in \cite{vs}.
Below we assume that $l =0$ (for $q\neq 1$ this corresponds to the fixed 
point reference frame but for $q=1$ this gives a nontrivial simplification).
The zero potential model considered in section 4 corresponds to a
particular solution of this system, $f_i(x)=\beta_j=const$, which 
determines thus the simplest self-similar potential.

Consider some examples. The $N=1$ case describes a $q$-deformation of the 
harmonic oscillator potential, since for $q= 1$ one has $f(x)=\mu x/2$ and 
$u(x)\propto x^2$. The $N=2, \; q=1$ system coincides with the 
conformal quantum mechanical model \cite{dealfaro},
\begin{equation}
f_{0,1}(x)=\pm{\gamma \over x}+\beta x, \quad
\gamma ={\mu_0-\mu_1\over 2(\mu_0+\mu_1)}, \qquad \beta={\mu_0+\mu_1\over 4},
\label{e7}  
\end{equation}
$$
u_{0,1}(x)= {\gamma (\gamma \pm 1) \over x^2}+\beta^2x^2 -\beta(1\mp 2\gamma )
+ \lambda_{0,1}.
$$
This is the  singular oscillator potential whose physical spectrum consists
formally of two arithmetic series but only one of them is physical 
due to the boundary condition at $x=0$. 
Coherent states of this model defined as eigenstates of the
symmetry operator $B^-$ were constructed by Barut and Girardello
\cite{barut} (for some amendments, see \cite{basu}). The $N=2, \; q=-1$
system obeying the same symmetry algebra 
will be considered in the next section.

It is natural to call the physics of the $N=2, \; q^2\neq 1$ models 
as the  $q$-deformed
conformal quantum mechanics because their symmetry algebra is $su_q(1,1)$.
A general solution of the basic equations (\ref{e15})
is not available in a closed form already for $N=1$.
Let us find $f_{0,1}(x)$  as formal series near $x=0$. 
Consider first singular solutions with the 
pole type singularity. Such solutions appear to be odd functions
with a simple pole at zero:
$$
f_0(x)={a\over x} +\sum_{i=1}^\infty b_i x^{2i-1},\qquad 
f_1(x)=-\; {a\over x} +\sum_{i=1}^\infty c_i x^{2i-1},$$
\begin{equation}
b_i+c_i=\sum_{j=1}^{i-1} {c_jc_{i-j}-b_jb_{i-j}\over 2i-1+2a},\quad
q^{2i}b_i+c_i=\sum_{j=1}^{i-1} {q^{2i}b_jb_{i-j}-c_jc_{i-j}\over 2i-1-2a},
\label{e25a}
\end{equation}
where $i=2, 3, \dots,$ and $a$ is an arbitrary parameter; the coefficients
$b_1, \; c_1$ have the form 
$$b_1={1\over 1-q^2}\left({\mu_0\over 1+2a}-{\mu_1\over 1-2a}\right),\quad
c_1={1\over 1-q^2}\left({\mu_1\over 1-2a}-{q^2\mu_0\over 1+2a}\right).
$$
In general, the series diverge at $q\to 1$, but for the special
choice of $a$ one gets the truncated solution (\ref{e7}). 
In the limit $q\to 0$, the function
$qf_0(qx)$ does not vanish, $qf_0(qx)\to a/x$, and then the potentials $u_{0,1}$
are obtained by simple dressing of the potential $a(a+1)/x^2$.
It is natural to expect that for $0 <q<1$ there exist $\mu_{0,1}$ such
that the series converge for arbitrarily large $x$. 
The condition $a(a+1)\geq 3/4$ guarantees that  
normalizable wave functions and their first derivatives vanish at zero 
\cite{dealfaro}. The spectrum of such a system would arise from only one 
geometric series (for the same reason that 
there is only one arithmetic series for the singular oscillator).

Due to the presence of the singularity one can restrict the space
to be a half-line, $0 < x <\infty$,  and interpret $x$ as
a radial coordinate appearing from the separation of variables in a
three-dimensional Schr\"odinger equation. The constant $a$ aquires then an
interpretation of orbital momentum of a particle when it takes integer
values $a=l$. Then in addition to the standard term $l(l+1)/x^2$ there is a very
complicated dependence of the potential on the quantum number $l$, indicating 
that in this picture the ``$q$-deformed Laplacian" is a nonlocal operator.

For the solutions that are non-singular at zero one has 
$f_{0,1}=\sum_{i=0}^\infty b_i^{(0,1)} x^i,$ 
where $b_0^{(0,1)}$ are two arbitrary constants. Again, in general the series 
diverge at $q\to 1$. A particular choice of initial conditions gives 
the solution which in this limit corresponds to (\ref{e7})
with a coordinate shift. 
In the $0<q<1$ region, the solution that is nonsingular for all real $x$
defines an infinite soliton potential whose spectrum is composed from 
two independent lowest weight irreducible representations of $su_q(1,1)$. 
In the limit $q\to 0$ it shrinks to the smooth two-soliton potential.
It is this solution that reduces to the $q$-oscillator one
(\ref{e11})-(\ref{e14}) (with $q$ replaced by $q^{1/2}$) 
after the restrictions $f_1(x)=q^{1/2}f_0(q^{1/2}x),\; \mu_1=q\mu_0$.
Probably without such a 
restriction this solution does not have a $q\to 1$ limit. In any case, at
$q\to 1$ the spectral series become equidistant, which means that the
potentials start to be unbounded at infinity. Obviously a more
complete and rigorous analysis of the structure of the solutions of the 
$N=2$ equations is necessary.

When $q=1$, already the $N=3$ case corresponds to transcendental potentials, 
namely, the $f_j(x)$ depend now on solutions of the Painlev\'e-IV (PIV)
equation \cite{vs}:
\begin{equation}
f_0(x)={1\over 2}\omega x+f(x),\qquad \omega = \mu_0+\mu_1+\mu_2,
\label{e88}
\end{equation}
$$
f_{1,2}(x)=-\; {1\over 2} f(x)\mp \frac{1}{2f(x)}(f^\prime(x)+\mu_1), 
$$
\begin{equation}
f^{\prime\prime}={{f^\prime}^2\over 2f}+{3\over2}f^3+2\omega x f^2+
\left(\mbox{$\frac{1}{2}$}\omega^2 x^2+\mu_2-\mu_0\right)f-{\mu_1^2\over 2f}.
\label{e99}
\end{equation} 
Thus, a simple generalization of the harmonic oscillator, characterized by the
split of its linear spectrum onto $N$ independent terms, is connected
with highly nontrivial ordinary differential equations
whose solutions are transcendental over the solutions of linear differential
equations with coefficients given by the rational and algebraic functions.
For the $N=3, \; q\neq 1$ system it is not possible to reduce the
order of the equation (i.e. no first integral is known). Because in this
case there are
solutions of (\ref{e15}) reducing in the limit $q\to 1$ to the 
PIV functions, it is natural to refer to this system
as to the $q$-deformed PIV equation.

The notion of $q$-periodicity (\ref{e15}) and the corresponding algebraic
relations  are central in this section.
Suppose that superpotentials $f_j(x)$ are not singular and that the operator 
$B^-$ is well defined ($B^\pm=(B^\mp)^\dagger$)
and has $N$ normalizable zero modes (the necessary condition for this 
is $E_k<E_{k+1}$),
\begin{equation}
B^-|l\rangle=0, \qquad \langle l | m\rangle=\delta_{lm}, 
\quad l,m=0, 1, \dots, N-1.
\label{vacua}\end{equation}
Then $|l\rangle$ represent the first $N$ bound states of 
self-similar potentials and the whole discrete spectrum
consists of $N$ independent geometric series:
$$
H|n\rangle=E_n|n\rangle, \qquad E_{n+N}=q^2E_n, \quad E_n< E_{n+1}< 0, 
$$
\begin{equation}
E_{kN+l}=E_l q^{2k}, \quad l=0, 1, \dots, N-1, \quad k=0, 1, \dots, \infty,
\label{energies}\end{equation}
$$
|Nk+l\rangle=C_k (B^+)^k |l\rangle, \qquad 
\langle Nk+l | Nk'+l'\rangle=\delta_{kk'}\delta_{ll'}.
$$
The normalization constants $C_k$ are determined up to an arbitrary
energy-dependent phase factor
$$
|C_k|^{-2}=\nu\prod_{s=0}^{N-1}\prod_{m=1}^k(1-q^{2m}E_l/E_s), \qquad
\nu=(-1)^N E_0E_1\dots E_{N-1}>0,
$$
and similar freedom exists in the action of raising and lowering operators
$$
B^+|kN+l\rangle = \prod_{s=0}^{N-1}\sqrt{E_lq^{2(k+1)}-E_s}|(k+1)N+l\rangle,
$$
\begin{equation}
B^-|kN+l\rangle = \prod_{s=0}^{N-1}\sqrt{E_lq^{2k}-E_s} |(k-1)N+l\rangle.
\label{action}\end{equation}
In the ``crystal" limit $q\to 0$ there remain only 
first $N$ levels and we get the general $N$-soliton potentials.

There may be  intermediate situations, when only some of the 
zero modes of $B^-$ satisfy necessary boundary conditions (the singular 
oscillator model is a known example). Then some of the spectral terms 
disappear; we do not consider such possibilities in detail here.
Note that for $q>1$ the above formulas probably only have formal meaning
in the differential Schr\"odinger equation case due to a too rapid growth 
of spectrum. Analysis of the $N=1$ potential showed that, for $q>1$, 
it has singularities, and thus the symmetry operators are not well
defined \cite{s1}. The finite-difference
Schr\"odinger operators, or the Jacobi matrices have a more rich
spectral structure \cite{askey,svz,spirizhed}. In particular, the spectrum
may consist of a discrete set of points only and be compact; it may grow 
exponentially fast, and the geometric series may accumulate near 
zero both from below and from above, etc. 

Let us summarize what one can do with the help of Darboux 
transformations. Starting from the zero potential $u(x)=0$, and 
performing $N$ dressing transformations, one can get $N$-soliton 
potentials. Starting from the latter and performing an infinite 
number of dressing transformations in 
the particular self-similar manner, we get a subclass of infinite 
soliton systems. Taking the $q\to 1$ limit we derive the 
Painlev\'e-type potentials, some of which degenerate into the finite-gap
potentials in the limit $\omega\to 0$. 
All these complicated systems are thus equivalent in 
some sense to the free Schr\"odinger equation.

Consider coherent states of the algebra 
(\ref{goodalgebra}) defined as eigenstates of the operator $B^-$, 
\begin{equation}
B^-|\alpha\rangle= \alpha^N|\alpha\rangle,
\label{deflarge}\end{equation}
where for convenience we denote the eigenvalue as the $N$th power of $\alpha$.
Since $B^-$ has $N$ linearly independent zero modes, there are 
$N$ independent states:
\begin{equation}
|\alpha_l\rangle =C_l(\alpha)\sum_{k=0}^\infty 
{\alpha^{kN+l}|kN+l\rangle\over \prod_{s=0}^{N-1}\prod_{m=1}^k
(E_lq^{2m}-E_s)^{1/2}},
\label{cohlarge}\end{equation}
$$
|C_l(\alpha)|^{-2} = \sum_{k=0}^\infty {|\alpha|^{2(kN+l)}\over 
\prod_{s=0}^{N-1}\prod_{m=1}^k (E_lq^{2m}-E_s) }
$$
$$
= |\alpha|^{2l} {_N\varphi_{N-1}}\left({0, \dots, 0 \atop b_1^l, 
\dots, b_{N-1}^l};
q^2, z\right), \quad z= |\alpha|^{2N}/\nu.
$$
$$
b_1^l= E_lq^2/E_0,\; \dots,\; b_{l+1}^l=E_lq^2/E_{l+1},\; \dots,\;
 b_{N-1}^l=E_lq^2/E_{N-1}.
$$
These coherent states are normalizable for $|\alpha|^{2N}<\nu.$
In order to find coherent states corresponding to the limit $q\to 1$
it is necessary to replace the $q$-products by the Pochhammer symbols.
The spectrum of the resulting  system is found from the formula 
$E_{n+N}=E_n+\omega$. The hierarchy of potentials with such a linear 
spectrum was investigated in \cite{vs}. Representation theory of the $q=1$
algebras was  considered in \cite{smith,rocek}. Their possible physical
applications were discussed also in \cite{kuz,dek,karas}.
In the limit $E_l \to E_0+l\omega/N, \; l=0, 1, \dots, N-1,$ one
finds the harmonic oscillator such that the states
$|\alpha_l\rangle$ constructed above coincide with the generalizations of 
even and odd coherent states (\ref{rootstates}).

The positive energy states are described by the following representation
\begin{eqnarray}
H\;|\lambda\rangle &=& \lambda\;|\lambda\rangle,  \quad \lambda>0,
\nonumber \\
B^+\;|\lambda\rangle &=& 
\prod_{s=0}^{N-1}\sqrt{\lambda q^2-E_s} |\lambda q^2\rangle,
\nonumber \\
B^-\;|\lambda\rangle &=& 
\prod_{s=0}^{N-1}\sqrt{\lambda -E_s} |\lambda q^{-2}\rangle.
\label{contgen}
\end{eqnarray}
When the states generated by $B^\pm$ from $|\lambda\rangle$ for some
fixed $\lambda$ are the only eigenstates of the Hamiltonian, 
generalization of the coherent states (\ref{newqcoh}) has the following form:
\begin{eqnarray}
|\alpha\rangle_\lambda= C(\alpha)\Biggl(
\sum_{n=0}^\infty {\alpha^{nN} q^{Nn(n-1)/2}\over
\lambda^{N/2}\prod_{s=0}^{N-1}\sqrt{(E_s/\lambda; q^2)_n}}
|\lambda q^{-2n}\rangle 
\label{cohcont1} \\
+\sum_{n=1}^\infty { \nu^{n/2}\over \alpha^{nN}} 
\prod_{s=0}^{N-1}\sqrt{(\lambda q^2/E_s; q^2)_n}
|\lambda q^{2n}\rangle \Biggr), \nonumber
\end{eqnarray}
where 
$$
|C(\alpha)|^{-2}= \sum_{n=0}^\infty {(-1)^{Nn} q^{Nn(n-1)} z^n
\over \prod_{s=0}^{N-1}(E_s/\lambda; q^2)_n} 
+ \sum_{n=1}^\infty {\nu^n \over |\alpha|^{2nN}} 
\prod_{s=0}^{N-1}(\lambda q^2/E_s; q^2)_n
$$
$$
= {_0\psi_N}(b_0, \dots, b_{N-1}; q^2, z), \quad 
z=(-1)^N|\alpha|^{2N}/\lambda^N, \quad b_l=E_l/\lambda.
$$
These states are normalizable for $|\alpha|^{2N}>\nu$.

Analogously, one can find the countable set of coherent states in the
case when positive energies form the  continuous spectrum: 
\begin{equation}
|\alpha\rangle_s = C(\alpha)
\int_0^\infty {d\lambda\; \lambda^{d_s}\;|\lambda\rangle
\over \prod_{l=0}^{N-1} \sqrt{(\lambda q^2/E_l; q^2)_\infty}},
\qquad d_s={\ln \sqrt{\nu}/\alpha^N q^2+2\pi i s\over \ln q^2},
\label{contlarge}\end{equation}
where the normalization constant $C(\alpha)$ is given in (\ref{normconst}) with
$\beta^2_l=-E_l$.
Zero modes of the Hamiltonian $H$ may be eigenstates of the operators
$A$ and $A^\dagger$ in the same way as in the $q$-Weyl algebra case;
there is no need to  describe them again. 

We conclude that coherent states of the self-similar
potentials are related to the basic hypergeometric series
$_N\varphi_{N-1}(z)$, $_0\psi_N(z)$ or to some integrals over
the latter functions. However, we do not know the explicit form of these
coherent states because the Hamiltonian eigenstates involve new complicated
transcendental functions whose complete analytical structure is not
accessible at present. Let us mention in passing that the finite-difference
analogs of some of the above self-similar potentials associated with the 
discrete Schr\"odinger equation have been described in \cite{svz}.
The discrete dressing chain in this case happens to define a discrete
time Toda lattice \cite{spirizhed}. 
A $q$-deformed supersymmetric interpretation of the models considered
in this and the preceeding two sections was given in \cite{s4,s2,and}.

\section{Two particular examples}
\setcounter{equation} 0

Consider in more detail the $N=2, \; q=-1,$ and a special subcase of
the $N=3, \; q=1$ closures of the dressing chain. In the first case, the 
system of equations determining  superpotentials has the form:
\begin{equation}
{d\over dx}(f_0(x)+f_1(x)) +f_0^2(x)-f_1^2(x)=\mu_0, 
\label{n2}
\end{equation}
$$
{d\over dx}(f_1(x)-f_0(-x)) +f_1^2(x)-f_0^2(-x)=\mu_1,
$$
where we assume that $\mu_0+\mu_1 \neq 0$. 
From the algebraic point of view this case is related to the $su(1,1)$ algebra
which serves as the formal spectrum generating algebra. If one assumes that
$f_0(x)$ (or $f_1(x)$) is antisymmetric, $f_0(-x)=-f_0(x),$ then our system
is equivalent to the $N=2,\; q=1$ case, or to
the singular harmonic oscillator model (\ref{e7}).
The coordinate region of this problem is restricted to
the half-axis which forbids the parity operator.
Indeed, the wave functions are not single-valued near $x=0$
for noninteger values of the parameter $\gamma$ and the action of the 
parity operator is not well defined. 
As a result,  the eigenvalue equation for the lowering operator 
$B^-=P(d^2/dx^2+\dots)$ is also not well defined (for integer $\gamma$, half of
the wave functions are singular, the coherent states have fixed parity, and the
action of $B^-$ and $B^-P$ may differ only by sign). 

It is easy to see that $f_0(x)$ or $f_1(x)$ cannot be symmetric, therefore 
the non-trivial solutions of (\ref{n2}) do not have fixed parity. 
In spite of its simplicity, the system (\ref{n2}) is hard to solve.
Let us represent $f_j(x)$ as sums of symmetric and antisymmetric parts:
\begin{equation}
f_{s,a}(x)=\frac{1}{2}(f_0(x)\pm f_0(-x)), 
\qquad g_{s,a}(x)=\frac{1}{2}(f_1(x)\pm f_1(-x)),
\end{equation}
and substitute this splitting into (\ref{n2}). Then it is not
difficult to find from these equations and their $x\to -x$ partners 
two ``integrals"
\begin{equation}
f_a(x)+g_a(x)=\sigma x, \qquad \sigma=\frac{1}{2}(\mu_0 + \mu_1 ),
\end{equation}
and
\begin{equation}
f_s^2(x)+f_a^2(x)=g_s^2(x)+g_a^2(x)+\tau,\qquad \tau=\frac{1}{2}(\mu_0-\mu_1).
\end{equation}
As a result, it is possible to express $f_a$ and $g_a$ via their 
symmetric partners
$$
f_a(x)={1\over 2\sigma x}\left( g_s^2-f_s^2+\sigma^2 x^2
+\tau\right),
$$
\begin{equation}
g_a(x)={1\over 2\sigma x} \left(f_s^2- g_s^2+\sigma^2 x^2-\tau\right).
\end{equation}
Eventually the original system of functional-differential 
equations is reduced to the form:
\begin{equation}
f_s'(x)=2g_s(x)g_a(x), \qquad g_s'(x)= -2f_s(x)f_a(x). 
\label{n2q-1}\end{equation}
This system of ordinary differential equations is related to
a Painlev\'e-V equation, 
which follows from the fact that we are considering
a subcase of the $N=4, \; q=1$ closure analyzed in \cite{vadler}.

For the special choice of initial conditions
$f_a(0)=g_a(0)=0$ one has $f_s^2(0)=g_s^2(0)+\tau$. Then the
singularity at $x=0$ cancels and one obtains a formal nonsingular solution 
of the equations (\ref{n2q-1}). It is possible that this solution
defines a smooth potential growing indefinitely when $|x|\to\infty$. 
Under this asumption, consequences of the presence of the symmetry 
algebra $su(1,1)$ in this model are different
from those for the singular oscillator potential. Namely,
the spectrum consists now of two arithmetic series, each being
determined by irreducible representations of the $su(1,1)$
algebra.  The coherent states $|\alpha_l\rangle$, $l=0, 1$,
or even and odd coherent states, are both physical 
but they are not eigenstates of the parity operator because the potential is 
not symmetric.

Another example that we would like to discuss in more detail
concerns the $N=3, \; q=1$ closure considered in \cite{vs}
(note that in this paper all Hamiltonians $L$, admitting $N$th order
differential operator $B$ as a symmetry operator satisfying the relation
$[L, B]=B$, were characterized). 
The PIV function appearing in this context has an infinite
set of rational solutions emerging for the specific choices of the
parameters $\mu_0, \; \mu_1, \; \mu_2$ \cite{luk}. Consider the following
rational solution:
\begin{eqnarray}
f_0(x)&=&{x\over 2}+{2x\over x^2+1}, \nonumber \\
f_1(x)&=&{x\over 2}-{1\over x}, 
\label{luksol}  \\
f_2(x)&=&{x\over 2}+{1\over x} -{2x\over x^2+1}, \nonumber
\end{eqnarray}
which corresponds to the constants $\mu_0=4,\; \mu_1=1,\; \mu_2=-2$.
The symmetry algebra has thus the form
$$
[L, B^\pm]=\pm 3B^\pm, \qquad 
$$
\begin{equation}
B^+B^-=L(L-4)(L-5), \qquad 
\end{equation}
$$
B^-B^+=(L+3)(L-1)(L-2), 
$$
where the Hamiltonian $L$ is 
\begin{equation}
L=-\; {d^2\over dx^2} +{4\over x^2+1}-{8\over (x^2+1)^2}+ \frac{1}{4} x^2
+\frac{3}{2}.
\label{hamdek}
\end{equation}
There are three normalizable eigenstates of the lowering operator $B^-$
corresponding to the lowest energy state, and second and third excited levels.
Acting by $B^+$ upon these three ``vacua" one finds the whole spectrum.
It consists of three arithmetic series forming the 
sequence $0, 3, 4, 5, 6, \dots$, i.e. there is a hole between 0 and 3
after which the spectrum becomes equidistant (the first excited state
with the energy 3 is generated by $B^+$: $|1\rangle\propto B^+|0\rangle$).

The explicit form of the coherent states, 
$B^-|\alpha\rangle=\alpha^3|\alpha\rangle$, is determined from the equation
\begin{equation}
\chi'''(x)-{6x\over x^2+1}\chi''(x)+{12x^2 \over (x^2+1)^2}\chi'(x)=
\alpha^3\chi(x),
\label{exam1}
\end{equation}
where
$$ \chi(x)\equiv (x^2+1)e^{x^2/4}\langle x | \alpha\rangle.$$
The author does not know any simple solution of this equation.
Each of the three linearly independent solutions defines a particular 
coherent state. Evidently, they have some common origin with the 
cubic root-of-unity superpositions of the coherent states (\ref{rootstates}).

The structure of spectrum hints that there are some additional 
symmetries in this model, and, indeed, there is a different set of 
raising and lowering operators satisfying the different algebra
found earlier in \cite{dek}. It corresponds to a different
solution of the dressing chain for the same Hamiltonian (\ref{hamdek}):
\begin{equation}
f_0(x)=-f_2(x)={x\over 2}+{2x\over x^2+1}, \qquad f_1(x)={x\over 2}, 
\label{soldek}
\end{equation}
for which $\mu_0=3, \; \mu_1=-2, \; \mu_2=0$. Now the symmetry algebra
has the form
$$
[L, B^\pm]=\pm B^\pm, \qquad
$$
\begin{equation}
B^+B^-=L(L-3)(L-1), 
\end{equation}
$$
B^-B^+=(L+1)(L-2)L.
$$
The operator $B^-$ has only two normalizable zero modes corresponding
to the first two levels of the Hamiltonian. The principally important feature
of this realization is that the lowest energy state is also a zero mode
of the raising operator. As a result, it is the first excited
level which serves as the lowest weight vector of the discrete
series representation of the above algebra from which the
raising operator $B^+$ creates an infinite tower of states. In this
picture the spectrum consists of only one arithmetic series
with step 1, and the isolated zero energy state forms a 
trivial one-dimensional irreducible representation of the algebra.
One may thus conclude that the spectrum 
generating algebra of a given potential, and hence the coherent states,
may be non-unique.

The explicit form of the coherent states of the second symmetry algebra are 
determined by the equation 
\begin{equation}
\chi'''(x)-x\left(1+{6\over x^2+1}\right)\chi''(x)
+\left(2+{8 \over x^2+1}-{12\over(x^2+1)^2}\right)\chi'(x)= \alpha^3\chi(x),
\label{exam2}
\end{equation}
where $\chi(x)$ is defined in the same way as in the first case.
Simple solutions of this equation are also not known to the author 
\cite{bron}. There should be only one physically acceptable
solution because now the coherent states have a unique expansion over the
Hamiltonian eigenstates for $\alpha\neq 0$
(i.e. there is no connection with the root-of-unity superpositions). 
An analogous situation has been described recently for a different 
particular Hamiltonian with equidistant spectrum \cite{fer}. 
It would be interesting to find physical characteristics of the coherent 
states (\ref{exam1}), (\ref{exam2}) and similar ones. 

\section{Schr\"odinger operators with discrete scaling symmetry}
\setcounter{equation} 0

Spectral theory of the one-dimensional Schr\"odinger equation (\ref{a})
for the bounded periodic potentials,
\begin{equation}
u(x+ l)=u(x),
\label{periodic}
\end{equation}
is well known \cite{ince}. 
The Bloch, or Floquet theorem states that there is at least one 
wave function $\psi(x)$ which can be represented in the form:
\begin{equation}
\psi(x)=e^{ikx}\phi(x),
\label{bloch}
\end{equation}
where $\phi(x)$ is a periodic function, $\phi(x+ l)=\phi(x)$. 
When $\psi(x)$ is a bounded function it defines a physically 
acceptable state. This requirement restricts the quasimomentum $k$
to be real. 

Consider a class of Schr\"odinger operators, which satisfy the relations
\begin{equation}
T^\dagger H=q^2HT^\dagger,\qquad H T = q^2 T H, \qquad 
T^\dagger T=T T^\dagger=1,
\label{weylagain}
\end{equation}
where $T$ is the unitary fixed affine transformation operator (\ref{afftr}).
The algebra (\ref{weylagain}) corresponds to the degenerate case of 
(\ref{goodalgebra}) 
when the polynomials of $H$ on the r.h.s. are replaced by a constant.
It is equivalent to the following constraint upon the potential:
\begin{equation} 
q^2u(qx+l )=u(x), 
\label{constraint}
\end{equation} 
which for $q=1$ is the periodicity condition (\ref{periodic}). Note however
that the limit $q\to 1$ is not defined uniquely. After the shift 
$u(x)\to u(x)+\sigma/(1-q^2)$ the condition (\ref{constraint}) is reduced 
in this limit to $u(x+l)=u(x)+\sigma$, which shows that $u(x)$ is the
sum of a periodic and Airy potentials. Only the additional requirement
$\sigma\to 0$ leads to the standard Bloch-Floquet theory.

For the shifted potential $v(x)=u(x+x^*)$, where $x^*$ is the fixed point,
$qx^*+l=x^*,$ the constraint (\ref{constraint})
is converted into $q^2v(qx)=v(x)$. This means that
\begin{equation}
v(x)=h(x)/x^2, \qquad h(qx)=h(x),
\label{qfloquet}
\end{equation}
where $h(x)$ is an arbitrary function obeying the indicated property
of periodicity on the logarithmic scale. In the following we restrict
ourselves to the case when $h(x)$ is bounded for $0< x<\infty$,
which is a natural coordinate region of the problem. In this case the 
potential $v(x)$ is singular at $x=0$ and vanishes in the limit 
$x\to \infty$. The value of $h(x)$ is not defined at zero unless it is a 
constant. This means that one cannot soften the singularity by requiring 
$h(x)$ to vanish at $x=0$. 
The formula (\ref{qfloquet}) defines a class of
potentials whose spectral theory we propose to call the $q$-Floquet theory.

It is instructive to rewrite the resulting Schr\"odinger equation using the
change of variables $x=\exp y$ and renormalization of the wave function
$\psi(x)=\sqrt{x}\chi(y)$:
\begin{equation}
- {d^2 \chi(y)\over dy^2} +(h(y)+\frac{1}{4})\chi(y)=\lambda e^{2y}\chi(y),
\label{qfloquety}\end{equation}
where $h(y+\ln q)=h(y)$ is now a periodic function.
This equation has a form similar to the original one but the normalization 
condition for bound state wave functions $\chi(y),$ 
\begin{equation}
\int_0^\infty|\psi(x)|^2 dx = \int_{-\infty}^\infty e^{2y}|\chi(y)|^2 dy =1,
\label{measure}\end{equation}
contains a nontrivial factor $e^{2y}$ in the measure, which diverges when
$y\to\infty$.

The presence of the non-trivial symmetry constrains the
structure of the wave functions.
The operator $T$ acts upon the wave functions as follows 
\begin{equation}
T\psi_i(x, \lambda)=\sqrt{q}\psi_i(qx+l, \lambda)
=\sum_{j=1}^2 c_{ij}\psi_j(x, q^2\lambda),
\label{basiceq}
\end{equation}
where $\psi_i(x, \lambda)$ are two linearly independent solutions
of the Schr\"odinger equation. 
On the one hand, this  $q$-difference equation shows that if 
$\psi(x,\lambda)$ is normalizable for some fixed $\lambda=\lambda_0$, 
then $T$ does not change its 
normalizability. On the other hand, the resulting wave function has 
the eigenvalue $\lambda=q^2\lambda_0$, which thus also belongs to the
physical spectrum. Applying the $T^{-1}$ operator one gets a
wave function with the eigenvalue $\lambda=q^{-2}\lambda_0$. This 
simple argument shows that if our Hamiltonian has physical states
with negative energy, then it exhibits the ``fall
onto the center" phenomenon \cite{landau,case}: the energy is not bounded 
from below. 

The class (\ref{qfloquet}) unifies periodic potentials 
with the conformal one corresponding to the special choice 
$h(x)=const$ \cite{dealfaro}. Indeed, the Hamiltonian $H=-d^2/dx^2+ h/x^2$, 
enters the following conformal symmetry agebra: 
\begin{equation}
[H,D]=iH, \qquad [K,D]=-iK, \qquad [H,K]= 2iD,
\label{conf}
\end{equation}
where $D=i\{x, d/dx\}/4$ and $K=x^2/4$ are 
the hermitian scaling and special conformal transformation generators. 
In this case dilatation by arbitrary parameter maps solutions of the 
Schr\"odinger equation
onto the solutions, which means formally that the spectrum of $H$ is purely 
continuous (such a situation holds only for $h>-1/4$). 
If $h(x)\neq const$ the discrete spectrum consists of a
number of geometric series infinite in both directions, 
with $\lambda=0$ and $\lambda=-\infty$ as the accumulation points. 
Note that (\ref{qfloquet}), with some unknown $h(x)$, was suggested as
the asymptotic form of the self-similar potential with one 
geometric series of bound states truncated from below \cite{degas}. 
In the more general case one has $N$ such series \cite{s2}; it is a matter of 
conjecture that in all cases asymptotics of the potentials have a similar form. 

From the symmetry point of view we have the following situation.
The simple conformal potential has a large group of continuous
symmetries which fixes the dynamics completely. We can
add an interaction term which does not remove all symmetries,
but preserves a discrete part of them. This situation is similar to 
the case where a free particle is put into a periodic potential --
instead of the group of continuous translations one has only a
discrete subgroup of it. 
Note that the generator of the discrete subgroup 
is now considered as an element of the symmetry algebra characterizing the 
spectrum -- this is one of the ways of building quantum algebras 
out of Lie algebras.

Due to the non-analyticity of the potential at $x=0$, the accurate spectral 
analysis  requires some rigorous mathematical tools. The main problem
is to find restrictions upon $h(x)$ for which the
Hamiltonian is self-adjoint, or for which it has self-adjoint extensions.
Here we give only a qualitative picture. From (\ref{qfloquet}) we see that 
there are three essentially different 
regions of the spectrum.  Positive energy states form continuous spectrum
because the potential goes to zero sufficiently fast.
Degeneracy of these states depends on the boundary condition
imposed upon the wave functions at $x=0$. Only for boundary conditions 
that are invariant under the taken scaling transformation, does
the operator $T$  represent a physical symmetry.

The second region of energy corresponds to zero modes of the Hamiltonian.
It is especially interesting because upon these zero modes $T$ commutes 
formally with the Hamiltonian. For $\lambda=0$, (\ref{qfloquety}) 
looks like the standard Schr\"odinger equation for a periodic potential
with spectral parameter equal to $-1/4$: 
$$
-\chi''(y)+h(y)\chi(y)=-\; \frac{1}{4} \chi(y), \qquad h(y+\ln q)=h(y).
$$
If both independent solutions of this equations are bounded, i.e. if the
eigenvalue $-1/4$ belongs to the permitted band of $h(y)$, then the original 
wave function $\psi(x)$ does not describe physical states, being unbounded at 
infinity. For boundedness of $\psi(x)$ at $x=\infty$
the quasimomentum $k$ of the Bloch eigenfunction 
$\chi(y)=e^{iky}\phi(y)$, $\phi(y+\ln q)=\phi(y)$, 
has to have an imaginary part Im $k\geq 1/2$
(for Im $k=1/2$ one has $\psi(qx)=q^{iRe\;k}\psi(x)$). Normalizability of 
$\psi(x)$ near $x=0$ imposes the restriction 
Im $k < 1$. For Im $k=1$, wave functions $\psi(x)$ are generalized 
eigenfunctions of $T$ with eigenvalues of modulus 1, as it should be for 
a unitary operator. The zero energy states of the Hamiltonian $H$ may
thus belong to the continuous spectrum when the quasimomentum $k$ of the 
Bloch function $\chi(y)$ is complex and $1/2 \leq {\rm Im}\; k \leq 1$.

For any $\lambda<0$ there
is a wave function $\psi(x)$ which is normalizable at infinity, so that
it is the boundary condition at $x=0$ which determines the quantization 
of the spectral parameter. If $h(x)>0$ then there are no bound states at all.
When squares of the absolute values of both solutions of the Schr\"odinger 
equation are integrable near zero, the Hamiltonian is not self-adjoint.
Its self-adjoint extensions are fixed by the requirement of a particular
dependence of wave functions near zero. 

Let us consider the case $h(x)=const$, which  exhibits  already basic 
features of the $q$-Floquet theory.
This model was considered by Case \cite{case} who has shown that
when $h<-1/4$ one can choose an orthonormal basis of states such
that the energies of bound states form an infinite 
geometric series in both directions. Such a spectrum is stipulated by the 
presence in the model of a discrete scaling symmetry. More precisely, the 
operator $D$ in (\ref{conf}) is not a physical symmetry operator any more, 
but the group element $q^{2iD}$ is -- it maps wave functions of the
discrete spectrum onto each other. Let us consider in more detail how this
situation appears using the Wronskian technique of self-adjoint
extensions of singular Schr\"odinger operators 
\cite{krall}. According to this approach, one takes
two solutions of the original equation 
for some fixed eigenvalue satisfying 
$\phi_1=\cos\alpha,$ $\phi_1'=\sin\alpha$ and
$\phi_2=\sin\alpha,$ $\phi_2'= -\cos\alpha$ at some regular point $x=\delta$
(evidently, $\phi_1\phi_2'-\phi_1'\phi_2=-1$).
Then it is necessary to take arbitrary linear combination
$\phi(x)=A\phi_1(x)+B\phi_2(x)$, $A, B$ real constants, 
and look for solutions of the Schr\"odinger equation whose Wronskian 
with $\phi(x)$ vanishes for $x\to 0$:
\begin{equation}
W(\phi, \psi)\equiv\phi(x)\psi'(x)-\phi'(x)\psi(x)\to 0, \qquad x\to 0.
\label{wronskian}
\end{equation}
Physically this means that the particle's behavior 
near the point of singularity
is fixed by the choice of auxiliary function $\phi(x)$.
When $\phi(x)$ and its derivative take finite values 
at $x=0$, the constraint (\ref{wronskian}) becomes equivalent 
to the well known condition $\psi'(0) = c\psi(0)$, where $c$ is a real constant.

For $h<-1/4$ derivatives of all wave functions are singular 
at $x=0$, and one needs to use a limiting procedure.
Let us take as auxiliary function the general zero energy eigenfunction
$$
\phi(x)= a\sqrt{x}\cos (\sigma \ln x +\theta),\qquad 
\sigma=\sqrt{|h|-1/4},
$$
where $a$ and $\theta$ are some real constants.
Substituting it into the above relation one finds,
$$
{d\over d\ln x} {\psi(x)\over\sqrt{x}} 
+ \sigma\tan (\sigma\ln x +\theta) {\psi(x)\over\sqrt{x}}\to 0,\quad x\to 0.
$$
The chosen auxiliary function is homogeneous under the scaling 
of coordinate by a specific constant $q=e^{\pi/\sigma}$, 
$\phi(qx)=-\sqrt{q}\phi(x)$. 
Therefore, from the scaled form of the condition (\ref{wronskian})
$$ 
\phi(qx)\psi'(qx)-\phi'(qx)\psi(qx) \propto \phi(x){d\psi(qx)\over dx} - 
{d\phi(x)\over dx}\psi(qx)\to 0, 
$$
it is seen that if $\psi(x)$ satisfies (\ref{wronskian}), 
the same holds for the scaled
wave function $\psi(qx)$, i.e. scaling by $q$ is a physical
symmetry of the problem.  This fact guarantees that
discrete eigenvalues appear in the form of a geometric series.
Since the behavior of the general solution of the Schr\"odinger equation
for small $x$ is known, the Wronskian vanishes only if  near zero 
$\psi(x)$ has the form of $\phi(x)$ taken
with the same angle $\theta$, which is a free parameter of the 
self-adjoint extension. The spectrum  itself is found from the requirement 
of normalizability of wave functions with such a property. 
The described technique of fixing the 
boundary condition at zero seems to be valid in the arbitrary case. 

As it was remarked in \cite{s2}, eigenfunctions of the Schr\"odinger
operators with self-similar potentials resemble discrete wavelets
\cite{meyer} -- the functions $\psi(x)$ affine transformations of which, 
$\psi_{j, n}(x)=2^{j/2}\psi(2^jx+n),$ generate an orthonormal basis of 
Hilbert space $L^2(${\bf R}), 
$\langle \psi_{j, n}|\psi_{l, m}\rangle=\delta_{jl}\delta_{nm}.$ 
In the above $q$-Floquet theory, we have a partial
realization of this construction because the scaling of the coordinate of
the wave functions by $q$ creates an orthogonal function. 
The connection with wavelets originates, of course, from
the group of affine transformations, because wavelets can be interpreted as
coherent states associated with unitary representations of this group 
\cite{as} according to one of the definitions mentioned in the introduction. 
An interesting fact that differential-delay equations similar to those
determining coherent states of self-similar potentials  may have solutions with 
finite support, or atomic solutions \cite{rvachev} 
indicates also a hidden relation with wavelets.

\section{The factorization method and ``quantum Galois theory"}
\setcounter{equation} 0

Let us discuss the problem of integrability of a given
equation. If this equation is algebraic, $P(x)=\sum_{j=0}^na_j x^j=0$,
with coefficients $a_j$ from some fixed number field $k$, then its 
solvability in terms of the radicals is determined by the solvability of 
its Galois group (the group of permutations of roots of $P(x)$) \cite{post}. 
One can say that algebraic equation is exactly solvable when the latter 
situation takes place, but this criterion is sensitive to the choice of $k$
-- there exists a universal field of complex numbers for which any algebraic 
equation is exactly solvable. 

Integrability, or exact solvability of differential
equations is a looser notion. The weakest definition requires the
existence of a solution of an initial value problem analytical in a 
bounded domain. The class of equations integrable in this sense is too large --
the field of functions determined by formal power series can be thought as a 
universal one since solutions of any differential equation with coefficients 
from this field are given by functions of the same type.
Another extreme definition consists in demanding for solutions to belong
to differential fields of elementary or classical special functions,
i.e. the fields built from the simplest functions and their derivatives.
The question of integrability acquires precise meaning when the 
functions entering the equations and those allowed in the solutions are 
completely specified. In short, the problem is to understand transcendence
of solutions over the given field of functions.

In fact there exists a natural analog of the 
Galois theory for fields with operators \cite{kap,singer}, which
can be considered physically as a ``quantum" (first-quantized) Galois theory.
The main notions of it are the Picard-Vessiot (P-V) extension
and the differential Galois group. 
Let $u(x)$ in $y''(x)=u(x)y(x)$ belong to the differential field 
$k=${\bf C}$(x)$ (field of rational functions). 
Roughly, the P-V extension $M=k\langle y_1, y_2
\rangle$ is built from rational functions of independent solutions
of this equation and their derivatives with coefficients from $k$. 
The group of automorphisms
of $k\langle y_1, y_2\rangle$ which commute with the derivative and keep 
$k$ fixed is called the Galois group $Gal( M/k)$. 
For a general $n$th order linear ODE it is isomorphic
to the group of invertable matrices $GL(n, ${\bf C}).
Roughly, the equation is solvable in terms of Liouvillian functions
(exponentials, integrals, algebraics of the functions from $k$)
when $Gal(M/k)$ is solvable (or more simply
when all elements of it can be made triangular).
This theory is not widely known although it forms
the basis of the intensively used computer programs of symbolic integration
of differential equations \cite{bronstein}.

Let $D$ be an indeterminate over some number field, then permutations of roots
of a polynomial of $D$ preserve this polynomial
$(D-x_1)\dots (D-x_n)=0 \to (D-x_{i_1})\dots (D-x_{i_n})=0$. 
In the quantum picture, when $D$ is an operator, e.g. $d/dx$, 
there are two analogs of these 
``classical" permutations -- the automorphisms figuring in the 
definition of the Galois group, which preserve the differential equation, and 
the permutations $f_i(x) \leftrightarrow f_j(x)$ in
$(D-f_1(x))\dots (D-f_n(x))y(x)=0$. The latter do not, in general, preserve 
the initial equation, but cyclic permutations of such type play a key role 
in the factorization method \cite{infeld,fordy}, which thus should be
considered as one of the ingredients of ``quantum" Galois theory. 

Consider the free Schr\"odinger equation with zero spectral parameter 
\begin{equation}
y''\equiv \left(D+ 1/x\right)\left(D-1/x\right)y=0
\label{zeroeig}\end{equation}
over the field of complex numbers {\bf C}. 
The P-V extension coincides  with {\bf C}$(x)$ and 
the Galois group consists of nontrivial triangular matrices.
Permutation of operator factors in (\ref{zeroeig}) gives the equation 
\begin{equation}
\left(D- 1/x\right)\left(D+ 1/x\right)y = 
y''-(2/x^2)y=0,
\label{perm}\end{equation}
which is different from the initial one, e.g. the potential
lies now in {\bf C}$(x)$. Its Galois group is trivial (consisting of the unit
matrix), because the general solution $y=a/x+bx^2$
belongs to {\bf C}$(x)$. In general, the  Darboux transformation 
changes the field of coefficients of ODE and its Galois group.
However, an important
feature is preserved in the chosen example; namely, solutions of both equations 
belong to one and the same field. If one factorizes the second 
equation using its general solution and permutes the operator factors, 
the solutions of the third equation, 
\begin{equation}
\left(D^2+{6x(2t-x^3)\over (t+x^3)^2}\right)y(x)=0,
\label{permtwo}\end{equation}
belong to {\bf C}$(x)$ again.
This procedure of building 
rational potentials out of the zero one with the help of 
Darboux transformations with zero eigenvalue level has been considered in
\cite{burch,lag,adler}. It can be shown that solutions of all equations built
in this way lie in the P-V extension of (\ref{zeroeig}),
i.e. the corresponding Galois groups are always trivial (an analogous situation
takes place for repeated Darboux transformations with non-zero eigenvalue 
level). A similar iso-Galois picture prevails, e.g. for the Bessel functions.
It would be interesting to investigate from this point of view the finite-gap
potentials for which one has a simple factorization of the Hamiltonian.

It is not difficult to describe a differential field which is preserved
by any Darboux transformation. 
For this, consider spectral problems generated by two 
Hamiltonians neighboring in the factorization chain 
\begin{equation}
(D\pm f(x))(D\mp f(x))y_{1, 2}(x)=\lambda y_{1, 2}(x).
\label{two}\end{equation}
Let $y(x, \lambda)$ be a particular solution of the first equation. Then 
\begin{equation}
y_1(x, \lambda)=y(x, \lambda)\left(a_1 + b_1\int^x {dx'\over y^2(x', \lambda)}
\right)
\label{twosol}\end{equation}
is a general solution.
Consequently, $y_2(x, \lambda)\equiv(D-f(x))y_1(x, \lambda)$ gives a 
general solution of the second equation for $\lambda\neq 0$. 
For $\lambda=0$ one has
\begin{equation}
y_2(x, 0)={1\over y(x, 0)}\left(a_2 +b_2\int^x y^2(x', 0) dx' \right).
\label{sols}\end{equation}
This function belongs to the P-V extension of the first equation
when the derivative $d y(x, \lambda)/d\lambda$ at $\lambda=0$ does so.
Indeed, one can verify by direct differentiation that 
\begin{equation}
\int^x y^2(x', \lambda)dx'=
y^2(x, \lambda){d^2 \ln y(x, \lambda)\over dxd\lambda}+ const.
\label{int}\end{equation}
This relation shows 
that Darboux transformations preserve the differential field 
generated by solutions of a spectral equation and their derivatives over 
$\lambda$ for arbitrary values of $\lambda$.
This is a very wide field, but it exists and may be used 
in a search for such potentials, $u(x)$, and values of $\lambda$
that $y(x, \lambda)$ are Liouvillian over the field from which $u(x)$ is taken.
It seems that the differential Galois theory, 
in conjunction with the equivalence up to a change of variables, 
can be used for a constructive definition of the 
notion of self-similarity. Indeed, 
for $u(x)=0$ and fixed $\lambda$ the P-V extension
contains only $\exp(\pm n\sqrt{\lambda} x) $, $n\in ${\bf N},
and this hints immediately at the form of the spectrum,
$\lambda_n\propto \lambda n^2$,
for which Darboux transformations preserve this field.
Analogously, defining relations of the 
self-similar potentials resemble the requirement of preservation
of the P-V extension after $N$ Darboux transformations.

In a sense, the power of the factorization method
(or of Darboux transformations) stems from
its relation to the Galois theory. This explains also a large number of
its applications. For instance, we mention its effectiveness
in numerical calculations \cite{wil}, in approximation theory
\cite{chud}, in search of bispectral equations \cite{grun},
in the theory of orthogonal polynomials \cite{svz,spirizhed},
its connection with symplectic structures \cite{wilson}, etc.
In particular, Darboux transformations with zero eigenvalue
appeared to play an important role in the construction of equations
satisfying the Huygens principle \cite{lag,berves}.

In conclusion, let us summarize the criteria which allow to call us potentials
``exactly solvable". First, such potentials should belong to some 
sufficiently simple differential field $k$, e.g. to {\bf C}$(x)$
but not to the field of formal power series. This requirement assumes 
that a full analytical structure of potentials is known. Then one
can say that the Schr\"odinger equation is exactly solvable if 
the solutions satisfying the
taken boundary conditions (i.e. when $\lambda$ belongs to the spectrum)
are given by Liouvillian functions over $k$. 
This definition is tied to the differential Galois theory 
where one has a simple test of such solvability. Note that in this
case a change of boundary conditions may change the ``solvability" of 
the equation. 
A weaker definition of exact solvability refers to the availability of global 
structure of solutions \cite{its}, i.e. it demands knowledge of 
various asymptotics of solutions, which often may be found even if the solution 
is not Liouvillian. The latter requirement is natural for spectral problems, 
since in order to satisfy boundary conditions one should be able to connect 
solutions at various distances. In this respect, presently only classical 
special functions provide a completely satisfactory bank of information. 

The situation with self-similar potentials is instructive. It 
is difficult to specify the field of functions to which they belong
(in particular, to find analytic properties of $f_j(x)$), but
once it is done the normalizable wave functions are given by Liouvillian
functions. The discrete spectrum of these potentials corresponds thus 
to the exactly solvable problem in the differential Galois theory sense. 
Analysis of the continuous spectrum states, or of the solutions 
satisfying different boundary conditions, requires information
which is not available at present, i.e.
the rank of special functions for solutions of Schr\"odinger equations with 
self-similar potentials is not yet established. 

Coherent states provide a basis of the Hilbert space of
states of quantum particle different from that determined by the Hamiltonian
eigenstates. It would be interesting to understand their role 
from the differential Galois theory point of view. 
For the standard harmonic oscillator case the picture is simple. 
The second order differential (Schr\"odinger) equation is replaced
by a first order one whose solutions provide an overcomplete set
of Hilbert space vectors. Certainly this provides a simplification 
and ``minimization" of the problem -- the non-physical eigenfunctions
of the Hamiltonian are removed in this procedure. 
Such minimization of the types of functions relevant to the given 
physical problem is characteristic for some 
coherent states from the functional-analytic point of view. 
This is evident for coherent states defined as orbits of states 
generated by physical symmetry groups. 
The ladder operator approach does not obey such a property in general
because, starting from $N=2$, coherent states of the self-similar potentials 
are determined by differential or differential-delay equations
of higher order, which may contain nonphysical solutions as well. 
Moreover, for $q\neq 1$ solutions of the latter
equations contain some functional arbitrariness with respect to 
the Schr\"odinger equation.

\section{Conclusions}
\setcounter{equation} 0

The conclusions are short. The superpositions of coherent states in the 
abstract form (\ref{rootstates}), generalizing the even and odd states 
of \cite{dodonov}, are applicable to a very wide variety of systems. 
The parity coherent states (\ref{newcoh}), 
(\ref{generalparity}) are less universal -- their meaning and 
the way of derivation are strongly tied to the presence of the
parity symmetry. For any parity invariant potential 
admitting ladder operators $A^\dagger, A$, one can make the canonical 
transformation $A\to VA, \; A^\dagger\to A^\dagger V$, where $V$ is a unitary
operator linear in the parity operator (\ref{generalparityop}), 
which  does not change the algebra satisfied by these operators and 
the Hamiltonian.
This is the most simple physical observation of the present work. 
For the discussion of experimental implementations
of superpositions of coherent states we refer to the papers 
\cite{yurke,mec,buzek,sanders,green,gravi} and references therein.

Coherent states of the free particle, constructed in section 4, differ 
qualitatively from those of the harmonic oscillator. 
First, they are built from the continuous spectrum states, 
which does not allow us to form superpositions similar to (\ref{rootstates}). 
Second, their analytical properties are quite unusual, so that at present 
it is not known even whether they are complete or not. It would be 
interesting to investigate the physical characteristics of these states.

In many respects the results of this paper are not complete.
This is caused by the complicated structure
of self-similar potentials and their limiting cases, namely,
by the appearance of Painlev\'e transcendents and their $q$-analogs.
However the author believes that this is a temporary situation and later
many things will be simplified and, more importantly, the appropriate 
phenomenological applications of the coherent states for self-similar 
potentials considered in sections 5 and 6 will be found. 
Because of such expectations, the open problems indicated
in this paper are interesting primarily from the physical point of view.
Among the problems worth further investigation, we mention the detailed 
analysis of various limiting cases of the type discussed in section 7, 
the $q$-Floquet theory, connections with wavelets, etc. On the 
mathematical side, it is necessary to find general structure of $q$-special
functions behind the self-similar potentials and to understand in a more
appropriate way the differential Galois theory origin of the notion
of self-similarity.

\bigskip\bigskip 

\noindent
{\Large \bf Acknowledgments}
\medskip

The author is indebted to Yu.Berest, M.Bronstein, D.Fivel, A.Iserles,
A.Its, V.P.Karassiov, 
I.Lutzenko, V.I.Man'ko, $\;$ A.Mann, M.Rahman, $\;$ Y.Saint-Aubin, 
A.Shabat, $\;$ M.Singer, S.Skorik, 
E.C.G.Sudarshan, S.Suslov, L.Vinet, and A.Zhedanov 
for helpful discussions of various aspects of
integrable systems and coherent states.
This work is supported by  NSERC of Canada 
and by the Fonds FCAR of Qu\'ebec.


\begin{thebibliography}{120}
\bibitem{bron} As M.Bronstein kindly informed the author there are no rational
and algebraic function solutions of the equations for $\chi(x)$ for any 
nonzero value of $\alpha$.
\end{thebibliography}
\end{document}